\documentclass[12pt, letterpaper]{article}

\pdfoutput=1

\usepackage{amsmath}
\usepackage{amsfonts}
\usepackage{amssymb}
\usepackage[colorlinks,citecolor=blue,urlcolor=blue,bookmarks=false,hypertexnames=true]{hyperref}
\usepackage{cite}

\usepackage{graphicx, physics, subcaption}
\usepackage{verbatim }

\numberwithin{equation}{section}

\usepackage{color}

\usepackage{amsmath, amssymb, graphics, epsfig, graphicx}
\usepackage{epsf}
\usepackage{epstopdf}
\usepackage {amssymb}
\newcommand{\nc}{\newcommand}
\nc{\ba}{\begin{eqnarray}}
\nc{\ea}{\end{eqnarray}}

\newcommand{\calR}{{\cal{R}}}
\newcommand{\calP}{{\cal{P}}}

\def\bfk{{\bf k}}

\def\bfx{{\bf x}}

\nc{\cN}{ {\cal{N}} }

\begin{document}

\vspace{5mm}
\vspace{0.5cm}
\begin{center}

\def\thefootnote{\fnsymbol{footnote}}

{\bf\large Stochastic  non-attractor inflation}
\\[0.5cm]

{ 
Hassan Firouzjahi$^{1}\footnote{firouz@ipm.ir},$
Amin Nassiri-Rad$^{1}\footnote{amin.nassiriraad@ipm.ir}, $
Mahdiyar Noorbala$^{2,1}\footnote{mnoorbala@ut.ac.ir} $
}
\\[0.5cm]

{\small \textit{$^{1}$ School of Astronomy, Institute for Research in Fundamental Sciences (IPM) \\ P.~O.~Box 19395-5531, Tehran, Iran
}}\\
{\small \textit{$^{2}$ Department of Physics, University of Tehran, Iran, P.~O.~Box 14395-547
}}\\

\end{center}

\vspace{.8cm}

\hrule \vspace{0.3cm}


\begin{abstract}

We extend the formalism of stochastic inflation to the setup of non-attractor inflation 
with a  sound speed $c_s$. We obtain the Langevin equations for the superhorizon perturbations and  calculate the stochastic corrections to curvature perturbation power spectrum. It is  shown that the fractional stochastic corrections in mean number of e-folds and  power spectrum are at the order of power spectrum. We also calculate the boundary crossing and
the first hitting probabilities in a hypothetical  dS space with two boundaries in field space. Furthermore, the stochastic corrections in power spectrum in 
a setup akin to  eternal inflation  with large diffusion term  are calculated.

\end{abstract}
\vspace{0.5cm} \hrule
\def\thefootnote{\arabic{footnote}}
\setcounter{footnote}{0}
\newpage

\section{Introduction}
\label{sec:intro}

Currently inflation is the leading paradigm for the dynamics of  early Universe cosmology. While it is still at the phenomenological stage and a deeper theoretical understanding of the mechanism behind the dynamics of inflation is missing, but its basic predictions are well consistent with cosmological observations \cite{Akrami:2018odb, Ade:2015lrj}. Among the basic predictions of models of inflation are that  the primordial perturbations are nearly Gaussian, nearly adiabatic and nearly scale invariant. 

In its simplest realization, inflation is driven by a scalar field which slowly rolls on top of its nearly flat potential. Usually, it is assumed that the inflaton field has reached to its attractor phase in which the dynamics of the system is purely determined by the value of the field, $\phi(t)$ while the velocity of the field $\dot \phi(t)$ does not play roles.   In other words, while the general phase space of the inflaton field is 
two-dimensional spanned by the variables $\large(\phi(t), \dot \phi(t) \large)$, but the attractor phase is a one-dimensional  subset of the phase space in which the value of the field plays the role of the clock, determining the whole dynamics of the system. This is well justified with the basic picture in which inflation is insensitive to classical histories so the attractor phase is the generic outcome of the inflationary dynamics. 

The simplest model of non-attractor inflation is the so called ultra slow-roll (USR) inflation in which the potential is very flat in a finite range of the field  values so the kinetic energy falls off exponentially \cite{Kinney:2005vj, Namjoo:2012aa, Martin:2012pe, Motohashi:2014ppa, Pattison:2018bct}.    In USR model, the would-be decaying mode of curvature perturbation (which is discarded in slow-roll models) is actually the growing mode. As a result, the curvature perturbation is not frozen on super-horizon scales and it undergoes an exponential growth. This is the key phenomenon distinguishing non-attractor models from the single field slow-roll scenarios which brought interests into models of non-attractor inflation in recent years. In particular, models of non-attractor inflation are among the very few known examples in literature which violate  Maldacena's consistency condition \cite{Maldacena:2002vr}.  Based on Maldacena's consistency condition, one can rule out ``all" models of single field inflation if  local-type non-Gaussianity is observed  with the amplitude  $f_{NL} \gtrsim 1$. However, non-attractor models violate Maldacena's consistency condition exactly because the curvature perturbations are not frozen on super-horizon scales \cite{Namjoo:2012aa}. Technically speaking, while the non-attractor setup is still a single field model but it is not a single clock  setup which was implicitly assumed in deriving Maldacena's consistency condition. This is because during the non-attractor regime the phase space is two-dimensional determined jointly by $\large(\phi(t), \dot \phi(t) \large)$ and one can not use $\phi(t)$ solely as the clock to parametrize the evolution of the system. 

One drawback of the USR setup is that inflation does not end and the curvature perturbations grow indefinitely. To remedy these problems, one has to terminate the non-attractor phase say e. g. by a waterfall mechanism. In this picture,  the whole inflationary period has two stages. The first stage is the non-attractor phase which lasts for a few e-folds. This is followed by the second stage which  is in attractor phase yielding a long period of slow-roll inflation. In principle, the transition from non-attractor phase to attractor phase can have important effects on the final curvature perturbations \cite{Cai:2016ngx, Cai:2017bxr}.

In USR setup with a flat potential, one may worry that the stochastic quantum fluctuations of the inflaton field may build up and have non-trivial effects on curvature perturbation power spectrum. Indeed, there have been interests in literature on the possibility of generating large curvature perturbations in USR setup on a window of scales as seeds for primordial black hole  formation during inflation \cite{Pattison:2017mbe,  Ezquiaga:2018gbw}, see also   \cite{Biagetti:2018pjj, Ballesteros:2020sre, Ragavendra:2020sop}. In \cite{Firouzjahi:2018vet} we have studied the quantum diffusion effects associated with the inflaton quantum fluctuations in the USR setup. Using the stochastic $\delta N$ formalism the corrections in curvature perturbations power spectrum and bispectrum are studied systematically. It was shown that the stochastic effects are sub-leading in power spectrum and bispectrum. More specifically, the fractional corrections in power spectrum and bispectrum are at the order of power spectrum and therefore are negligible. In this paper we extend the analysis of \cite{Firouzjahi:2018vet} to the $P(X)$ model of non-attractor inflation  in which the inflaton field has non-standard kinetic energy. The $P(X) $ non-attractor inflation model was studied  in \cite{Chen:2013aj, Chen:2013eea} as  an extension of simple USR setup where the sound speed of the cosmological perturbations $c_s$ plays non-trivial roles in curvature perturbations and the amplitude of 
non-Gaussianity.

The rest of the paper is organized as follows. In section \ref{stoc-rev}  we present the formalism of stochastic inflation which will be used for the $P(X)$ setup presented in section 
\ref{P(X)-rev}.  In section \ref{power-sec} we calculate the mean number of $e$-folds and  the power spectrum using stochastic $\delta N$ formalism. In section  \ref{hitting} we calculate the first hitting probabilities and the  first boundary crossing for the classical motion and quantum jumps. In section \ref{large kappa} we consider the stochastic effects in the new limit of large diffusion term followed by summary and conclusions in section \ref{summary}.  Some technicalities of stochastic calculus and the sub-leading corrections of analysis in section \ref{large kappa} are relegated to  appendices \ref{calculus} and \ref{app:moment} respectively.


\section{ Stochastic Inflation}
\label{stoc-rev}

Here we briefly present  the formalism of stochastic inflation which we employ in models of inflation with non-standard kinetic terms. Stochastic inflation formalism for a field with a non standard  kinetic term was studied in the context of DBI inflation in \cite{Lorenz:2010vf}. 

In stochastic formalism, the quantum fluctuations of light scalar fields, such as the inflaton field, are decomposed into the long and short wavelengths perturbations. The small scale perturbations inside the Hubble horizon act as active source of noises for long mode perturbations outside the Hubble horizon. In a inflationary background with a near dS like background, these noises are Gaussian with the amplitude $H/2\pi$ in which $H$ is the Hubble expansion rate during inflation. For various works related to stochastic inflation see  \cite{Vilenkin:1983xp, Starobinsky:1986fx, Rey:1986zk, Nakao:1988yi, Sasaki:1987gy, Nambu:1987ef, Nambu:1988je,  Kandrup:1988sc,  Nambu:1989uf, Mollerach:1990zf, Linde:1993xx, Starobinsky:1994bd, Kunze:2006tu, Prokopec:2007ak, Prokopec:2008gw, Tsamis:2005hd, Enqvist:2008kt, 
Finelli:2008zg, Finelli:2010sh, Garbrecht:2013coa, Garbrecht:2014dca, Burgess:2014eoa,  Burgess:2015ajz, Boyanovsky:2015tba,  Boyanovsky:2015jen, Fujita:2017lfu, Talebian:2019opf, Talebian:2020drj, Pinol:2020cdp}. 

The stochastic formalism with long and short modes decomposition is a natural setup to employ $\delta N$ formalism. More specifically, the $\delta N$ formalism \cite{Sasaki:1995aw, Sasaki:1998ug, Lyth:2004gb, Wands:2000dp, Lyth:2005fi, Abolhasani:2019cqw, Abolhasani:2018gyz} is based on the separate Universe approach in which the super-horizon perturbations affect  the background expansion of the nearby patches (Universes). $\delta N$ formalism is  a powerful tool to calculate the curvature perturbation power spectrum and non-Gaussianity. The  extension of $\delta N$ formalism to stochastic inflation has been studied in  \cite{Fujita:2013cna, Fujita:2014tja, Vennin:2015hra, Vennin:2016wnk, Assadullahi:2016gkk, Pattison:2019hef, Grain:2017dqa, Noorbala:2018zlv}.

The model we are interested in enjoys a shift symmetry  so the action is a function of $X\equiv -\frac{1}{2} g^{\mu \nu} \partial_\mu \phi \partial_{\nu} \phi$ and there is no dependence on $\phi$. In particular, as in USR model, the potential  is flat, $V=V_0$. The USR setup is a particular example with $P=X$ which was studied in our previous work. The shift symmetry employed here is for simplification in order to handle the equations analytically. However, there is no limitation in employing stochastic formalism to general case in which the action is a function of $\phi $ with $P= P(X, \phi)$. Models with non-standard kinetic energy, such as DBI inflation, have been extensively studied in literature. One important prediction of these models is that large equilateral type non-Gaussianity can be generated with the amplitude of non-Gaussianity proportional to $1/c_s^2$ in which $c_s$ is the sound speed of scalar perturbations
\ba
\label{cs}
c_s^2 =\frac{P_{, X}}{P_{, X} + 2 X P_{, X X} } \, ,
\ea
where  $P_{,X}$ denotes a derivative with respect to $X$ and so on. 

The action is given by
\begin{equation}\label{1}
    S=\int d^4x\sqrt{-g}\Big(\frac{R}{2}+P(X)\Big) \, ,
\end{equation}
where $R$ is the Ricci scalar with the reduced Planck mass set to unity, $M_P=1$. 

Starting with the FLRW metric
\ba
ds^2 = -dt^2 + a(t)^2 d {\bf x}^2 \, ,
\ea
the scalar field equation is given by
\begin{equation}\label{2}
     P_{,X}\partial_{\mu}\partial^\mu\phi+P_{,XX}\partial_\mu X\partial^\mu\phi+3HP_{,X}\dot{\phi}=0 \, ,
\end{equation}
in which  a dot denotes the derivative with respect to $t$ and 
$H= \dot a(t)/a(t)$ is the Hubble expansion rate.

Following \cite{Nakao:1988yi, Sasaki:1987gy}, we split $\phi$ and its time derivative $v \equiv \dot \phi$ into the short and long wavelengths as follows 
\begin{equation}
\label{3}
\begin{split}
\phi\left(\bfx,t\right)=\phi_l\left(\bfx,t\right)+\sqrt{\hbar}\phi_s\left(\bfx,t\right),
\end{split}
\end{equation}
\begin{equation}
\label{4}
\begin{split}
v\left(\bfx,t\right)=v_l\left(\bfx,t\right)+\sqrt{\hbar}v_s\left(\bfx,t\right),
\end{split}
\end{equation}
in which the labels  $l$ and $s$ denote the long and short modes respectively.  To specify the quantum natures of the short modes we have inserted the factor 
 $\sqrt \hbar$  in Eqs. (\ref{3}) and (\ref{4}). With this decomposition, to leading order in $\sqrt \hbar$,  $X$ becomes 
\begin{equation}\label{5}
    X=\frac{v_l^2}{2}+v_lv_s\sqrt{\hbar} \, .
\end{equation}

Going to Fourier space , the short modes satisfy the following decomposition,   
\begin{equation}
\label{5}
    \phi_s\left(x,t\right)=\int\frac{d^3 \bfk}{\left(2\pi\right)^3}\theta\left(c_s k-\varepsilon aH\right)\phi_\bfk\left(t\right)e^{ik.x},
\end{equation}
and
\begin{equation}
\label{6}
    v_s\left(x,t\right)=\int\frac{d^3 \bfk}{\left(2\pi\right)^{3}}\theta\left(c_s k-\varepsilon aH\right)\dot\phi_\bfk\left(t\right)e^{ik.x} \,
\end{equation}
where $\theta$ is the step function. 
We have introduced the small dimensionless number $\varepsilon \ll 1$ to separate the large and small scales in an appropriate way ($\varepsilon $ should not be confused with the slow-roll parameter $\epsilon$). 
In addition, in this setup with a non-trivial $c_s$ the Hubble horizon is replaced by the sound horizon so $k$ is accompanied by a factor $c_s$.  In addition, $\phi_{\bfk }(t)$ is written in terms of the annihilation and creation operator as   $\phi_\bfk=a_\bfk\varphi_k+a^\dagger_{-\bfk}\varphi_{k}^*$ in which $\varphi_k$ is the positive frequency mode function of scalar perturbations.

Now we substitute Eqs. \eqref{3}  and \eqref{4} into \eqref{2} and expand the result up to first order in $\sqrt{\hbar}$, yielding  the following equations\cite{Nakao:1988yi, Sasaki:1987gy} in phase space
\begin{equation} \label{7}
    \dot \varphi_l =v_l+\sqrt{\hbar}\, \sigma,
\end{equation}  
and  
\begin{equation}\label{8}
     P_{,X}\dot{v}_l+P_{,XX}\dot{v}_lv_l^2=-3HP_{,X}v_l-(P_{,X}+P_{,XX}v_l)\sqrt{\hbar}\, \tau \, ,
\end{equation}
where we have neglected the spatial derivatives of $\phi_l$. In addition,   $\sigma$ and $\tau$ are the 
stochastic noises originating from the short modes which 
affect the evolution of the long modes. More specifically,  $\sigma$ and $\tau$ are related to quantum mode functions via  
\begin{equation}
    \label{9}
    \sigma\left(\bfx,t\right)= \varepsilon aH^2\int\frac{d^3 \bfk}{\left(2\pi\right)^3}\delta\left(c_s k-  \varepsilon aH\right)\phi_\bfk\left(t\right)e^{i \bfk \cdot \bfx},
\end{equation}
\begin{equation}
    \label{10}
    \tau\left(\bfx,t\right)= \varepsilon aH^2\int\frac{d^3 \bfk}{\left(2\pi\right)^3}\delta\left(c_s k-  \varepsilon aH\right)\dot\phi_\bfk\left(t\right)e^{i \bfk \cdot \bfx}.
\end{equation}

Our goals here are to obtain the Langevin equations for the long modes and to calculate the correlation functions of the noises.  Defining the mode functions $u_k\equiv z\mathcal{R}_k$ with 
$\mathcal{R}_k=\frac{-H}{\dot{\phi}}\varphi_k$, the equation of motion for the scalar perturbation is given by \cite{Chen:2006nt} 
\begin{equation}\label{10}
u''_k+(c_s^2k^2-\frac{z''}{z})u_k=0,
    \end{equation}
in which  $z\equiv \frac{a\sqrt{2\epsilon}}{c_s}$,  $\epsilon \equiv  -\dot H/H^2$ is the first slow-roll parameter and  a primes denotes a   derivative with respect to conformal time $d \tau = dt/a(t)$.

Imposing the Bunch-Davies (Minkowski) initial condition for the modes deep inside the horizon, the solution is given by 
\begin{equation}\label{12}
    \varphi_k=\frac{iHc_s}{2\sqrt{P_{,X}c_s^3k^3}}\left(1+ikc_s\tau\right)e^{-ikc_s\tau}.
\end{equation}
With the profile of wave function given above, we are able to obtain the amplitude of noise at the time of sound horizon crossing $c_s k= a$H. Compared to \cite{Nakao:1988yi, Sasaki:1987gy}, the wavefunction is multiplied by 
$c_s/\sqrt{P_{,X}}$ while $k\rightarrow c_sk$. Correspondingly, the amplitude of the noise is scaled via \cite{Lorenz:2010vf}
\ba
\frac{1}{c_s^4P_{,X}}\times{c_s^3}=\frac{1}{c_sP_{,X}}.
\ea
Consequently,  compared to \cite{Nakao:1988yi, Sasaki:1987gy},  the vacuum  correlation functions of $\sigma$ and $\tau$ are given as follows
\begin{equation}\label{sigmacor}
    \left<\sigma(x_1)\sigma(x_2)\right>=\frac{H^3}{4\pi^2c_sP_{,X}}j_0\Big(\varepsilon \frac{aH}{c_s}|x_1-x_2|\Big)\delta\left(t_1-t_2\right)
\end{equation}
\begin{equation}\label{taucor}
    \left<\tau(x_1)\tau(x_2)\right>=  \varepsilon^4 \frac{H^5}{4\pi^2c_sP_{,X}}j_0\Big(\varepsilon \frac{aH}{c_s}|x_1-x_2|\Big)\delta\left(t_1-t_2\right)
\end{equation}
\begin{equation}
    \left<\sigma(x_1)\tau(x_2)+\tau(x_2)\sigma(x_1)\right>=-2 \varepsilon^2 \frac{H^4}{4\pi^2c_sP_{,X}}j_0\Big(\varepsilon \frac{aH}{c_s}|x_1-x_2|\Big)\delta\left(t_1-t_2\right)\, ,
\end{equation}
where $j_0$ is the zeroth order spherical Bessel function.  Moreover, the commutator of the $\tau$ and $\sigma$ are given as follows
\begin{equation}
    \left[\sigma(x_1),\sigma(x_2)\right]=\left[\tau(x_1),\tau(x_2)\right]=0 \, ,
\end{equation}
\begin{equation}\label{commutator}
    \left[\sigma(x_1),\tau(x_2)\right]=i\varepsilon^3\frac{H^4}{4\pi^2c_s P_{,X}}j_0\Big(\varepsilon \frac{aH}{c_s}|x_1-x_2|\Big)\delta\big(t_1-t_2\big) \, .
\end{equation}

As we see from \eqref{commutator},   the commutator of $\sigma$ and $\tau$ goes to zero when $\varepsilon \rightarrow 0$  and hence the quantum nature of the noises disappear on superhorizon limit. 
In addition, from  Eq.  \eqref{taucor} we see   that  $\tau$  goes to zero on superhorizon scales while the  correlation for $\sigma$ simplifies to 
\begin{equation}
\left<\sigma(x_1)\sigma(x_2)\right>=\frac{H^3}{4\pi^2c_s P_{,X}}\delta(t_1-t_2)=\frac{H^4}{4\pi^2c_s P_{,X}}\delta(N_1-N_2),
\end{equation}
where in the last equality we have used the number of e-folds, defined via $d N=H dt$, as the clock.

With these considerations and noting that $\tau \rightarrow 0$, 
the Langevin equations for the superhorizon modes from  Eqs.  \eqref{7} and \eqref{8} are obtained to be 
\begin{equation}
\label{langevin}
   \frac{d\phi}{dN}=\frac{v}{H}+\frac{H}{2\pi\sqrt{c_s P_{,X}}}\xi(N),
\end{equation}
\begin{equation}
\label{vevolve}
\frac{dv}{dN}+3Hc_s^2v=0,
\end{equation}
where we have dropped the subscript $l$ and defined $\sigma\equiv\frac{H}{2\pi}\xi(N)$ in which $\xi(N)$ is a classical white noise with the following properties
\begin{equation}
\left<\xi(N)\right>=0,\quad \left<\xi(N)\xi(N')\right>=\delta(N-N').
\end{equation}

Equations (\ref{langevin}) and (\ref{vevolve}) are the starting points for our analysis 
in  next sections.

\section{$P(X)$ non-attractor inflation}
\label{P(X)-rev}

The analysis presented in previous section was general for any $P(X)$ model, independent of the system being in attractor or non-attractor phase.   In this section we employ the stochastic formalism outlined above to  non-attractor $P(X)$ model. As mentioned in Introduction, the $P(X)$ model is the extension of the USR setup which was studied in details in \cite{Chen:2013aj} and \cite{Chen:2013eea}, see also \cite{Abolhasani:2019cqw}. One motivation to study non-attractor $P(X)$ model was to obtain non-trivial contributions from $c_s$ in power spectrum and $f_{NL}$.  As discussed before, one drawback of the above setup (like the USR setup) is that inflation does not end. Indeed, as inflation proceeds the kinetic energy of the field falls off exponentially and the background approaches more and more to the dS spacetime. One can cure this problem by gluing the non-attractor phase to a follow up attractor phase. If the transition from non-attractor phase to the attractor phase is sharp, one may expect that the super-horizon perturbations may not be affected. However, in general, the transition may affect the final curvature perturbation power spectrum and bispectrum   \cite{Cai:2016ngx, Cai:2017bxr}.

While our results in this section are valid for any $P(X)$ non-attractor model, but to have a specific example, one can consider the following model \cite{Chen:2013aj, Chen:2013eea}, 
\begin{equation}
\label{18}
    P\left(X\right)=X+\beta X^\alpha-V_0 \, ,
\end{equation}
where  $\alpha$ and $\beta$  are positive constants.  The conventional USR setup corresponds to the case $\beta=0$.  Note that as in USR setup, the dominant source of background expansion is given by 
$V_0$ so to leading order in slow-roll parameter, $H^2 \simeq V_0/3$.  As discussed in \cite{Chen:2013aj, Chen:2013eea}, during the non-attractor phase the non-linear term containing $\beta$ dominates. In this limit, during the non-attractor phase, we obtain $c_s^2 \simeq 1/(2\alpha-1)$.
As the system reaches the attractor phase and the kinetic energy falls off significantly, the usual linear term $X$ takes over.

Now considering a general $P(X)$ non-attractor model and 
assuming a near constant $H$, which is a very good approximation in non-attractor setup,  
from  Eq. \eqref{vevolve} we obtain 
\begin{equation}\label{21}
    \dot{X} + 6 H c_s^2 X =0 \, ,
\end{equation}
where the relation $\dot X = v_l \dot v_l$ and the formula Eq. (\ref{cs}) for $c_s$ 
 have been used. 

Using the number of e-fold $N$ as the clock via $H d t = d N$, the above equation can be solved yielding 
\begin{equation}\label{22}
X\left(N\right) =  X_0\exp\left(-6Nc_s^2\right),
\end{equation}
where $X_0$ is the initial kinetic energy of the field at the start of the non-attractor phase $N=0$.
From the above equation we find that 
\begin{equation}\label{23}
    v_l=\sqrt{2X}=\sqrt{2X_0}\exp\left(-3Nc_s^2\right).
\end{equation}
Plugging this expression into Eq. \eqref{langevin} we obtain the following Langevin equation for $\phi(N)$
\begin{equation}\label{24}
    \frac{d\phi}{dN}=\frac{\sqrt{2X_0}}{H}\exp\left(-3Nc_s^2\right)+\frac{H}{2\pi\sqrt{c_s P_{,X}}} \, \xi\left(N\right),
\end{equation}
where, as mentioned before,  $\xi\left(N\right)$ is a white Gaussian noise.

The above equation can be integrated yielding
\begin{equation}
\label{25}
    \phi\left(N\right)= \phi_0 + \frac{\sqrt{2X_0}}{3Hc_s^2}\left(1-\exp\left(-3Nc_s^2\right)\right)+\frac{H}{2\pi \sqrt{c_s}}\int^N_0\frac{1}{\sqrt{P_{,X}}}\xi\left(N'\right)dN'.
\end{equation}
where $\phi_0$ is the initial value of the field. 

Now let us define the surface of end of inflation (or the end of non-attractor phase, whichever comes first ) when $\phi$ reaches a fixed value $\phi=\phi_e$. Classically and in the absence of the stochastic noise, this happens at $N=N_c$ 
which from Eq. (\ref{24}) is given by
\begin{equation}
    N_c = -\frac{1}{3c_s^2} \ln \left[ 1 - \frac{3Hc_s^2(\phi_e-\phi_0)}{\sqrt{2X_0}} \right] \, .
\end{equation}
However, in the presence of the stochastic noises, different patches take different number of e-folds before hitting the surface of end of inflation. As a result, the time of end of inflation, denoted by ${\cal N}$, is a stochastic parameter. Correspondingly, Eq. (\ref{25}), can be written as 
\begin{equation}\label{26}
    \exp\left(-3\mathcal{N}c_s^2\right)=\exp\left(-3N_cc_s^2\right) \Big[1+\kappa\int^{\mathcal{N}}_0\frac{1}{\sqrt{P_{,X}}}\xi\left(N'\right)dN' \Big],
\end{equation}
where we have defined the dimensionless parameter $\kappa $ via
\begin{equation}\label{kappa-def}
    \kappa\equiv\frac{3H^2c_s^2\exp\left(3N_c c_s^2\right)}{2\pi\sqrt{2X_0 c_s}} \, .
\end{equation}
Since $\kappa$ is a key parameter of our analysis, it is useful to express it in terms of observable quantities.
Indeed, using Eq. (\ref{22}) relating $X(N)$ to $X_0$ and the relation $\epsilon \equiv -\dot H/H^2 = X P_{,X}/H^2$ for the first slow-roll parameter, one can show that
\ba
\label{kappa-P}
\kappa = 3 c_s^2 \sqrt{P_{,X}(N_c)}  \left( \frac{H^2}{ 8 \pi^2 c_s \epsilon (N_c)} \right)^{1/2} = 
3  c_s^2 \sqrt{P_{,X}(N_c)} \sqrt{ \calP_\calR ^{(0)}(N_c)} \, ,
\ea
in which ${\calP_\calR ^{(0)}}(N_c)$ is the curvature perturbation power spectrum calculated at the end of non-attractor phase. The superscript $^{(0)}$ here means that it is the classical power spectrum, i.e. power spectrum in the absence of stochastic noises. On the physical ground we expect $\kappa \ll1$ so a perturbative expansion in powers of $\kappa $ is allowed. However, in Section~\ref{large kappa} a hypothetical case violating  $\kappa \ll 1$ is studied which would be the case for eternal inflation.

Now let us see how the small parameter $\kappa$ varies along a given classical background trajectory. In the classical limit (in the absence of noise), from Eq. (\ref{25}) the background field equation is given by  
\begin{equation}
\label{phi-c}
    \phi\left(N\right)=\phi_0+\frac{\dot\phi_0}{3Hc_s^2}\left(1-\exp\left(-3Nc_s^2\right)\right).
\end{equation}
From the above equation we see that each trajectory is characterized by a maximum field excursion 
$\phi_\text{max}$ given by
\ba
\label{phi-max}
\phi_\text{max} \equiv  \phi_0 + \frac{\dot\phi_0}{3Hc_s^2}\, ,
\ea
beyond which the field can not roll classically.    We can easily find
\begin{align}
    \dot\phi_0 = 3Hc_s^2 (\phi_\text{max} - \phi_e) e^{3N_cc_s^2}, \\
    \phi_0 = \phi_\text{max} - (\phi_\text{max} - \phi_e) e^{3N_cc_s^2}.
    \end{align}
Since both $\phi_\text{max}$ and $\phi_e$ are fixed on a given trajectory, these relations enable us to see how $\phi_0$ and $\dot\phi_0$ vary along the trajectory with changing $N_c$.  We can now infer from Eq.~\eqref{kappa-def} that $\kappa \propto \exp\left(3N_cc_s^2\right) / \dot\phi_0$ is independent of $N_c$.

Finally, we remind that the velocity of the field falls off exponentially during the non-attractor phase. Specifically, from Eq. (\ref{phi-c}) or simply from Eq. (\ref{22}) we have $X(N) = X_0 e^{- 6 c_s^2 N} $. This is the main reason why the curvature perturbations are not frozen on superhorizon scales, yielding to the violation of the non-Gaussianity consistency condition \cite{Namjoo:2012aa, Chen:2013aj, Chen:2013eea, various-consistency}.


\section{ Power Spectrum}
\label{power-sec}

In this section, using the stochastic $\delta N$ formalism, we calculate curvature perturbation power spectrum and the stochastic corrections.  A similar analysis for the simple USR setup with $P(X)=X$ and 
$c_s=1$ were performed in \cite{Firouzjahi:2018vet}.  

To use the stochastic $\delta N$ formalism, we have to calculate $\langle \cN \rangle $ and 
$\delta \cN^2  \equiv \big \langle (\cN - \langle \cN \rangle)^2 \big\rangle=  \langle \cN^2 \rangle - \langle \cN \rangle^2$ in which the curvature perturbation power spectrum 
is given by \cite{Fujita:2014tja, Vennin:2015hra, Firouzjahi:2018vet}
\begin{equation}
\label{37}
     \mathcal{P}_\calR=\frac{d\left\langle\delta\mathcal{N}^2\right\rangle}{d\left\langle\mathcal{N}\right\rangle} \, .
\end{equation}
 
To start with,  let us  take the logarithm of both sides of \eqref{26} and expand the  result in terms of 
$\kappa$,
\begin{equation}
\label{27}
    \mathcal{N}=N_c+\frac{1}{3c_s^2}\sum^{\infty}_{n=0}\frac{(-\kappa)^n}{n}  {\bf W}(\cN)^n \, ,
\end{equation}
where we have defined  
\ba
d W(N) \equiv \xi(N) dN  \, ,
\ea
and
\ba
\label{hatW}
   {\bf W}(\cN) \equiv \int_0^\cN \frac{d W(N)}{\sqrt{P_{,X}(N)}} \, .
\ea
Note that $W(\cN)$ is the Wiener process \cite{Evans} associated with the noise $\xi(N)$ satisfying 
\ba
\label{Weiner}
\langle W(\cN) \rangle =0 \, , \quad \quad  \langle W(\cN)^2 \rangle = \langle \cN \rangle \, .
\ea
Also note that the upper bound of the integral Eq. (\ref{hatW}), ${\cal N}$, is a stochastic quantity. Furthermore,   the non-linear factor $1/\sqrt{P_{,X}}$ in Eq. (\ref{hatW}) makes the relation between ${\bf W}(\cN)$ and $dW$ non-trivial. 

Taking the stochastic average of Eq. (\ref{27}), we have
\ba
\label{N-av}
\langle \cN \rangle = N_c + \frac{\kappa^2}{6 c_s^2} \langle {{\bf W}(\cN)}^2 \rangle
-\frac{\kappa^3}{9 c_s^2} \langle {{\bf W}(\cN)}^3 \rangle  +  \frac{\kappa^4}{12 c_s^2} \langle {{\bf W}(\cN)}^4 \rangle +O(\kappa^5)
\ea
Note that we are interested into the leading stochastic corrections in curvature perturbation power spectrum. As a result, we have to calculate $\langle \cN \rangle$ and $\delta \cN^2$ to $O(\kappa^4)$. For this purpose, we need $\langle {{\bf W}(\cN)}^2 \rangle$,  $\langle {{\bf W}(\cN)}^3 \rangle$ and $\langle {{\bf W}(\cN)}^4 \rangle$ to orders of $\kappa^2$, $\kappa$ and $\kappa^0$ respectively. 

Similarly, to calculate $\delta \cN^2$, from Eq. (\ref{N-av}) and (\ref{27})   we have
\ba
\delta \cN = \cN - \langle \cN \rangle = -\frac{\kappa}{3 c_s^2} {\bf W}(\cN)  +  
  \frac{\kappa^2}{6 c_s^2}  {{\bf W}(\cN)}^2 -  \frac{\kappa^3}{9 c_s^2}  {{\bf W}(\cN)}^3 +  \frac{\kappa^4}{12 c_s^2}  {{\bf W}(\cN)}^4 + O(\kappa^5)
\ea
yielding
\ba
\label{deltaN2}
\hspace{-0.5cm}\delta \cN^2 &= &\big \langle \cN^2 \big \rangle - \langle \cN \rangle^2 \nonumber\\
& =&  \frac{\kappa^2}{9 c_s^4} \langle {{\bf W}(\cN)}^2 \rangle
-\frac{\kappa^3}{9 c_s^4} \langle {{\bf W}(\cN)}^3 \rangle  +  \frac{\kappa^4}{108 c_s^4} \Big[11 \langle {{\bf W}(\cN)}^4 \rangle -3\langle {{\bf W}(\cN)}^2 \rangle^2\Big]  +O(\kappa^5)
\ea

To proceed further, we need to calculate various correlations involving $\langle {{\bf W}(\cN)}^n \rangle$. This 
is somewhat non-trivial which we present the details in Appendix \ref{calculus}. The key is to use the stochastic integrals  in which \cite{Evans}
\begin{equation}\label{29}
\begin{split}
&\big\langle \int^\mathcal{\mathcal{N}}_0f\left(N\right)dW\big\rangle=0 \, ,\\ 
&\big \langle \big(\int^\mathcal{\mathcal{N}}_0f\left(N\right)dW\big)^2\big \rangle=\big \langle \int^\mathcal{N}_0f\left(N\right)^2dN\big \rangle.
\end{split}
\end{equation}
Note that in the above integrals $f(N)$ is an arbitrary deterministic function but the upper bound of integral is a stochastic variable. 

Furthermore, using the Ito lemma Eq. (\ref{Ito}), we  can relate $\langle {{\bf W}(\cN)}^n \rangle$ to $\langle {{\bf W}(\cN)}^{n-2} \rangle$ via
\begin{equation}\label{45}
     \big \langle {{ {{\bf W}(\cN)}^n}} \big\rangle=\frac{n(n-1)}{2} \big \langle \int^\mathcal{N}_0 \frac{{\bf W}\left(N\right)^{n-2}}{P_{,X}}dN \big\rangle \, .
\end{equation}
In particular we have
\begin{equation}\label{45p}
     \big \langle{{{{\bf W}(\cN)}^3}}\big\rangle=3\big \langle \int^\mathcal{N}_0 \frac{{\bf W}\left(N\right)}{P_{,X}}  dN \big\rangle \, ,
\end{equation}
and
\begin{equation}\label{46}
   \big \langle{{ {{\bf W}(\cN)}^4}} \big\rangle=6\big \langle \int^\mathcal{N}_0 \frac{{\bf W}\left(N\right)^2}{P_{,X}}  dN \big\rangle  \, .
\end{equation}

Collecting the results, to leading orders in $\kappa$, we have (see Appendix \ref{calculus} for details ) 
\ba
 \big \langle{{{{\bf W}(\cN)}^2}}\big\rangle=
  \big( 1+     \frac{\kappa^2}{6 c_s^4 P_{,X}} \big)  { I}(N_c) + O(\kappa^4) \, ,
\ea
\ba
 \Big \langle{{{{\bf W}(\cN)}^3}}\Big\rangle= -\frac{\kappa}{c_s^2 P_{,X}}  { I}(N_c)   + O(\kappa^3) \, ,
\ea
and
\ba
\big \langle{{{{\bf W}(\cN)}^4}}\big\rangle= 6{ { I}(N_c)}^2
-6   \int^{N_c}_0 \frac{dN}{P_{,X}}    { I}(N)   + O(\kappa^2) \, ,
\ea
where we have defined the integral $ { I}(N)$ as 
\ba
\label{I-def}
{ I}(N) \equiv \int^{N}_0 \frac{dN'}{P_{,X}(N')}  \, .
\ea
Note that $I(N)$ is a non-stochastic function. 

Now we are in a position to calculate the leading stochastic corrections in various correlation functions, such as $\langle \cN \rangle$ and $\calP_\calR$. For the average number of e-folds, from Eq. (\ref{N-av}) we obtain
\begin{equation}
\label{28}
     \left\langle \mathcal{N}\right\rangle=N_c+\frac{\kappa^2}{6c_s^2}  { I}(N_c) +  \frac{\kappa^4}{36c_s^4 P_{,X}} 
     \left[ 5 { I}(N_c)  + 18 c_s^2 P_{,X} \Big(  { { I}(N_c)}^2 -   \int^{N_c}_0 \frac{dN}{P_{,X}}    { I}(N)  \Big) \right]
     +O\left(\kappa^6\right).
\end{equation}
and
\begin{equation}
\label{deltN-2}
\delta \cN^2 =  \frac{\kappa^2}{9c_s^4}  { I}(N_c) + \frac{\kappa^4}{108 c_s^8 P_{,X}} 
\left[ 2(1+ 6 c_s^2)  { I}(N_c) + c_s^4 P_{,X} \Big( 63 { { I}(N_c)}^2 - 66 \int^{N_c}_0 \frac{dN}{P_{,X}}    { I}(N)   \Big)
\right] +O\left(\kappa^6\right).
\end{equation}
From Eq. (\ref{28}) we see that the stochastic corrections in $\langle \mathcal{N}\rangle$ is at the order of the power spectrum ($\propto \kappa^2$) which is negligible. 

With $ \left\langle \mathcal{N}\right\rangle$ and $\delta \cN^2$ calculated above, and using the definition of $\kappa$ given in Eq. (\ref{kappa-P}), 
the curvature perturbation power spectrum from Eq. (\ref{37}) is obtained to be
\ba
\label{power-final}
\calP_\calR = \frac{d\left\langle\delta\mathcal{N}^2\right\rangle}{d\left\langle\mathcal{N}\right\rangle} &=&  
 \frac{d\left\langle\delta\mathcal{N}^2\right\rangle/dN_c}{d\left\langle\mathcal{N}\right\rangle/dN_c}
\nonumber\\ 
&=&\calP_\calR^{(0)}  \Big[ 1 +  \calP_\calR^{(0)}  \Big(\frac{3}{2} (1+ 5 c_s^2)+ 45 c_s^4 P_{,X} 
 { I}(N_c) \Big) \Big]  +O\left(\kappa^6\right).
\ea
The above formula is interesting, showing that the fractional stochastic corrections in power spectrum is at the order  of power spectrum. Hence,  the stochastic corrections in power spectrum is sub-leading.  

One can check that for the simple USR case with $P= X$ and $c_s=1$, Eq. (\ref{power-final}) reproduces the result in \cite{Firouzjahi:2018vet}. It is also interesting to apply the above general result to the particular 
$P(X)$ model Eq. (\ref{18}). In the limit discussed there, the leading order corrections to 
$\langle \cN \rangle $ is obtained to be
\ba
\langle \cN \rangle = N_c + \frac{ c_s^2 \Big( 1- e^{-3 (1- c_s^2) N_c}  \Big)}{2( 1- c_s^2)} \calP_\calR^{(0)} 
+ O(\kappa^4)
\ea
Similarly, the  curvature perturbation power spectrum is
\ba
\calP_\calR = \calP_\calR^{(0)}  \Big\{ 1 +  \calP_\calR^{(0)}  \Big[\frac{3}{2} (1+ 5 c_s^2)+ 
\frac{15 c_s^4}{1- c_s^2} \big( 1-  e^{-3 (1- c_s^2) N_c} \big) \Big]
 \Big\}  +O\left(\kappa^6\right).
\ea
In particular, for the simple USR case with $c_s=1$, the above equations yield to
\ba
\langle \cN \rangle = N_c \Big(  1 + \frac{3 }{2}  \calP_\calR^{(0)} \Big)   + O(\kappa^4) \quad \quad (USR)
\ea
and
\ba
\calP_\calR = \calP_\calR^{(0)} \Big(1+ 9 (1+ 5 N_c) \calP_\calR^{(0)}
 \Big) +O\left(\kappa^6\right) \quad \quad (USR)
\ea
in  agreement with the results of \cite{Firouzjahi:2018vet}.


\section{Boundary Crossing Probabilities}
\label{hitting}

Here we extend the analysis of \cite{Firouzjahi:2018vet} to our setup 
studying  a hypothetical question  in which there are two 
absorbing barriers in field space located at $\phi_+ > \phi_0$ and $\phi_- <\phi_0$. The field hits either of the barriers and inflation ends after that.  Our goal is to  calculate  the first boundary crossing probabilities $p_+$ and $p_-$ which are  the probabilities of hitting first either $\phi_+$ or $\phi_-$ respectively. This setup 
may not be relevant for the observable  inflationary period as it may take a large number of $e$-folds for the field to hit either barriers but nonetheless this is an interesting conceptual question. Based on our assumption, if  we wait long enough  the field hits either barriers so we have $p_+ + p_- =1$.

Let us define $\cN$ as the total number of $e$-folds required for the field $\phi$ to hit either of the barriers.
Also, because of the shift symmetry involved in our setup, we assume the field starts at the origin in field
space corresponding to $\phi_0 =0$. Using Eq. (\ref{25}), the equation for  $\phi(\cN)$ is 
\begin{equation}
\label{25b}
    \phi\left(\cN \right)=  \frac{\sqrt{2X_0}}{3Hc_s^2}\left(1-\exp\left(-3 \cN c_s^2\right)\right)+\frac{H}{2\pi \sqrt{c_s}} {\bf W}(\cN)  
\end{equation}
where ${\bf W}(\cN)$ is as defined in Eq. (\ref{hatW}). 

Since there are two absorbing barriers $\phi_\pm$, it is not easy to solve the above equation analytically
so one has to employ numerical analysis.  In special limits we solve the question of first boundary crossing analytically and compare them with the full numerical results.

\subsection{ Brownian motion}
\label{ssec:BrownianMotion}

One limit that the first boundary crossing can be solved analytically is when there is no drift ($X_0=0$) so we have a pure quantum noise. This corresponds to a pure Brownian motion in a dS spacetime where the noise arises from
the quantum fluctuations of the scalar field. Of course, this is a oversimplified case compared to slow-roll inflation where the field has a small classical drift, but  nonetheless it provides interesting  theoretical  insights. 

In this case, the evolution of fluctuations of $\phi$ is given by the following Brownian-like equation 
\begin{equation}
\label{25c}
    \phi\left(\cN \right)=  \frac{H}{2\pi \sqrt{c_s}} {\bf W}(\cN)   \, .
\end{equation}
Now, taking  the stochastic average  of the above equation  and noting that  
$ \langle {\bf W}(\cN) \rangle =0  $ we obtain 
\ba
\label{Brownian-phi}
\langle \phi(\cN) \rangle =0 \, .
\ea 
This is interesting indicating that on average the field is stuck to the origin. 

\begin{figure}[t!]
		\centering
		\includegraphics[scale=0.5]{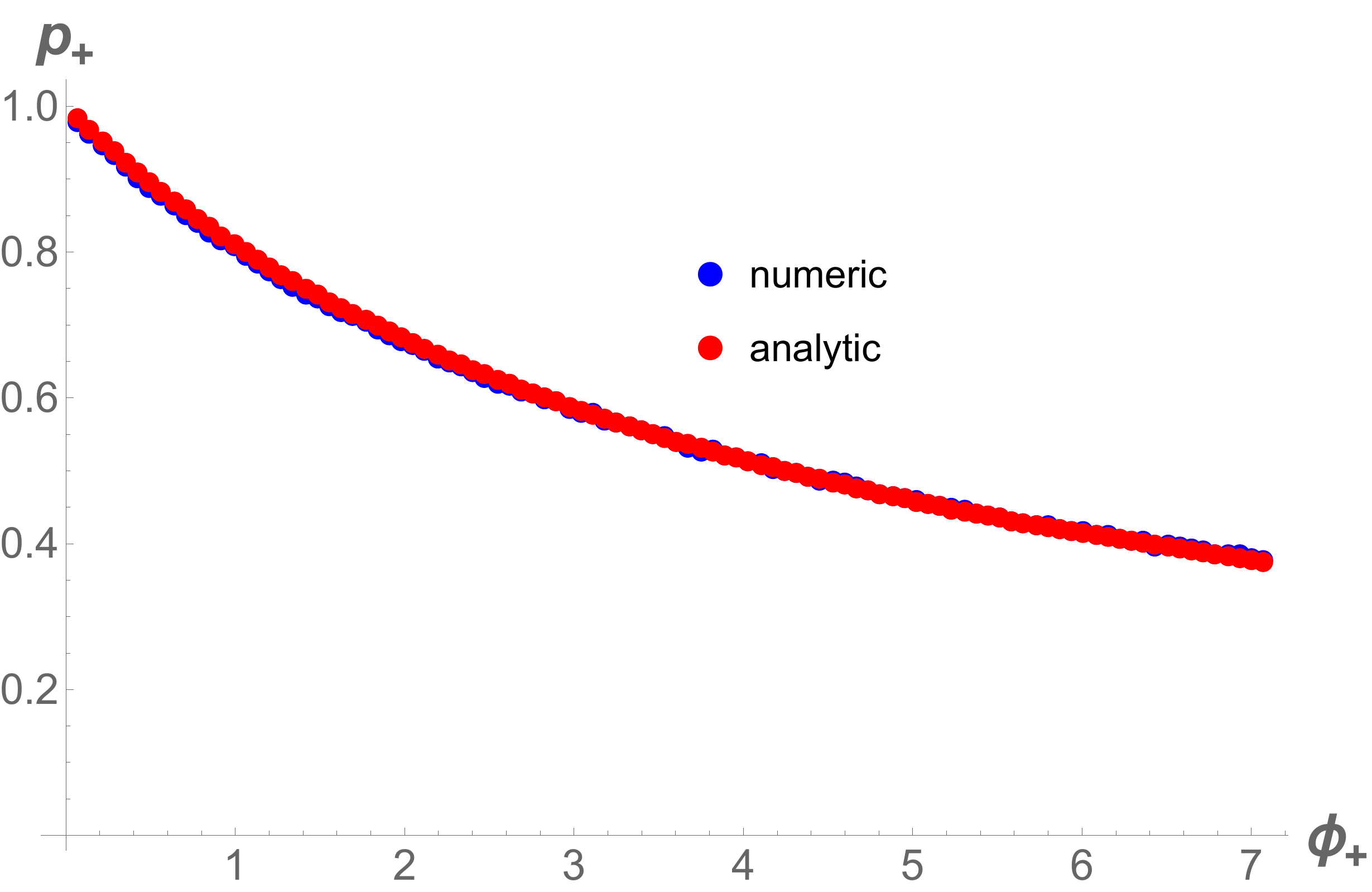}
		\hspace{0cm}
		\includegraphics[scale=0.5]{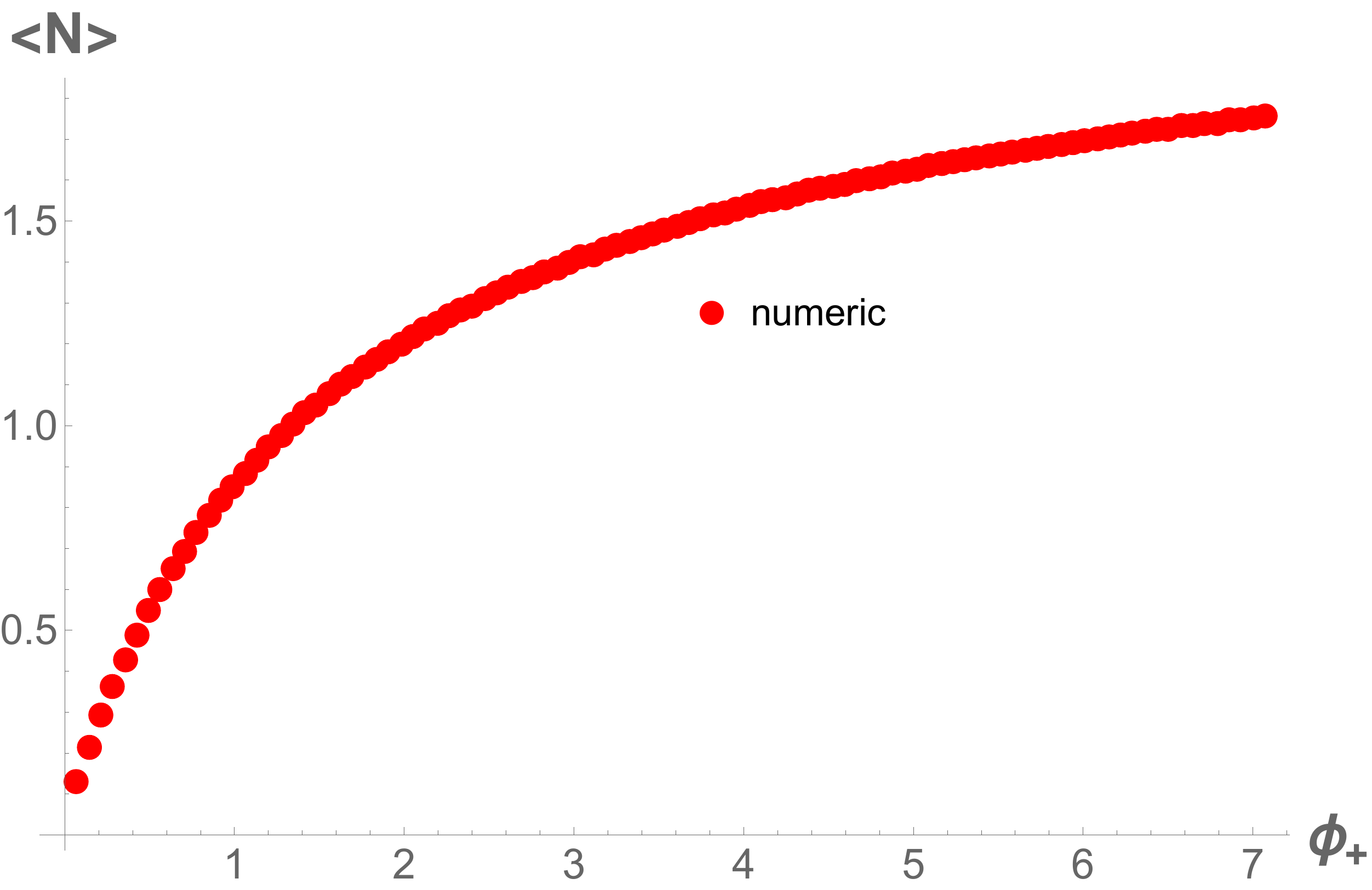}	
		\vspace{1cm}
	\caption{Top: the figure shows the simulated and theoretical results for $p_+$
	 in the case with no drift. The field values are normalized by $\frac{H}{2\pi\sqrt{c_s}}$ and we have fixed  $\phi_-=-4.2$ while  $\phi_+$ is changed from 0.07 to 7.  Bottom:  the simulated mean number of e-folds again in the case without drift. In both figures we have set $c_s= 0.5$ with $P(X)$ given in Eq. \eqref{18}. }
	\label{Brownian-plots}
\end{figure}

On the other hand, from the definition of $p_+$ and $p_-$ one calculates $\big \langle \phi(\cN) \big \rangle$  to be 
\ba
\label{phi-av-Brownian}
\big \langle \phi(\cN)  \big\rangle = p_+ \phi_+ + p_- \phi_- \, .
\ea
Combining  Eqs. (\ref{phi-av-Brownian})  and    (\ref{Brownian-phi})    we  obtain the following results for $p_\pm$, 
\ba
\label{Brownian-probs}
p_+ = \frac{-\phi_-}{\phi_+ - \phi_-} \, , \quad \quad
p_- = \frac{\phi_+}{\phi_+ - \phi_-} \, .
\ea
Interestingly, the above results for $p_\pm$ are exactly the same as in \cite{Firouzjahi:2018vet}. We conclude that  the first hitting probability for a given barrier  is proportional to the distance of the mirror barrier to the origin. The further away is the mirror barrier, the higher is the probability to  first hit the given barrier.  
If either of $\phi_+$ or $  \phi_-$ is pushed to infinity, then $p_+ $ or  $p_-$ goes to zero respectively 
which is understandable intuitively. The behaviour of $p_+$ is plotted in top panel of 
Fig.  \ref{Brownian-plots}. The agreements between the full numerical results obtained from simulations  and the theoretical values are excellent.

Having obtained $p_\pm$, the next step is  to calculate $\langle \cN \rangle$, the average number of e-folds 
when the filed hits either of  the barriers. Because of the non-trivial effects of $P(X)$ in ${\bf W}(\cN)$, this analysis is more difficult that the case of \cite{Firouzjahi:2018vet}. To start with, taking  the expectation value of the square of Eq. (\ref{25c}) we obtain  
\ba
\big \langle \phi(\cN)^2 \rangle &=&  \Big( \frac{H}{2\pi \sqrt{c_s}}  \Big)^2  \big\langle {\bf W}(\cN)^2 \big \rangle \, ,
\nonumber\\
&=& \Big( \frac{H}{2\pi \sqrt{c_s}}  \Big)^2  \big\langle I(\cN) \big \rangle \, ,
\ea
where $I(\cN)$ is defined in Eq. (\ref{I-def}). 

On the other hand, using the values of $p_\pm$ from Eq.  (\ref{Brownian-probs}) we have 
\ba
\label{phi2-av-Brownian}
\big \langle \phi(\cN)^2 \big \rangle = p_+ \phi_+^2 + p_- \phi_-^2  = - \phi_- \phi_+ \, .
\ea
Combining the above two equations, we obtain the following implicit equation 
\ba
\label{N-av-Brownian}
 \langle I(\cN) \rangle =  \Big(  \frac{\phi_+}{\frac{H}{2\pi \sqrt{c_s}} } \Big)  \Big( \frac{-\phi_-}{\frac{H}{2\pi \sqrt{c_s}} }  \Big) \, .
\ea
Note that the above equation is only an implicit equation for $\langle \cN \rangle$ as $I(\cN)$ is a non-linear function of $\cN$ and it is not straightforward to obtain $\langle \cN \rangle$ directly from Eq. (\ref{N-av-Brownian}). 

To interpret the meaning of Eq. (\ref{N-av-Brownian}) note that intuitively speaking 
$H/2\pi\sqrt{c_s}$   represents  the length of each quantum jump (with the effects of sound speed included)  so the ratios  $\phi_+/(H/2\pi\sqrt{c_s})$ and  $-\phi_-/(H/2\pi \sqrt{c_s})$ respectively measure the classical displacements of $\phi_+$ and $-\phi_-$ relative to  quantum jumps to reach either of the barriers. In the special case when the initial position of the field is located on the position of either barrier then 
$  I(\langle \cN  \rangle ) = \langle \cN \rangle =0$ and one of
$p_\pm$ is equal to unity while the other one is zero. For example, if we have $\phi_+=0$, then $p_+=1$ and $p_-=0$. 

Eq. (\ref{N-av-Brownian}) is valid for any theory of $P(X)$. Now as an example let us apply this to our special case of $P(X)$ given in Eq. (\ref{18}),  yielding
\ba
\frac{ \big \langle  e^{3 (1- c_s^2) \cN}  \big \rangle   -1 }{3 (1- c_s^2)}  =  \frac{- \phi_- \phi_+ }{ \big(\frac{H}{2\pi \sqrt{c_s}} \big)^2} \, .
\ea
Note the non-trivial combination $\big \langle  e^{3 (1- c_s^2) \cN}  \big \rangle$ which is non-linearly related to $\langle \cN \rangle$ and higher multipoles. 

In the limit $\cN \ll 1$, one can neglect the higher multipoles in $\big \langle  e^{3 (1- c_s^2) \cN}  \big \rangle$, yielding 
\ba
\label{N-av-Brownian-b}
 \langle \cN \rangle \simeq  
 \frac{- \phi_- \phi_+ }{ \big(\frac{H}{2\pi \sqrt{c_s}} \big)^2}
 \,   \quad \quad (\cN \ll 1)  \, .
\ea
This agrees with the result in \cite{Firouzjahi:2018vet} when $c_s=1$.  The linear dependence of  $\langle \cN \rangle $ to $\phi_+$ for a fixed value of $\phi_-$ can be seen in Fig.  \ref{Brownian-plots}
for small field values when $\langle \cN \rangle$ is also small. 

On the other hand, for large $\cN \gg1$, one may roughly take 
\ba
\label{largeN-a}
\big \langle  e^{3 (1- c_s^2) \cN}  \big \rangle \sim   e^{3 (1- c_s^2)  \langle \cN \rangle } \, ,
\ea
and obtain the following approximate result for $\langle \cN \rangle$
\ba
\label{largeN-b}
\langle \cN \rangle \sim \frac{1}{3 (1- c_s^2) } \ln \Big[ 3 (1- c_s^2)  \frac{- \phi_- \phi_+ }{ \big(\frac{H}{2\pi \sqrt{c_s}} \big)^2}
\Big] \, .
\ea
We emphasis that Eq. (\ref{largeN-b}) provides only a rough estimation of $\langle \cN \rangle$ as 
Eq. (\ref{largeN-a}) is valid only approximately.  

It is instructive to compare the behaviour of $\langle \cN \rangle$ in our case with the corresponding case 
in \cite{Firouzjahi:2018vet} when $c_s=1$. In that setup $I(\cN)$ is linear so $\langle \cN \rangle$ grows linearly in terms of $\phi_+$ for a fixed value of $\phi_-$. Here, however, $P(X)$ is non-linear so $I(\cN)$ becomes non-linear and the behaviour of $\langle \cN \rangle$ is very different than the linear dependence
as can be seen  in Fig. \ref{Brownian-plots}. Indeed, one sees a logarithmic dependence as is suggested from our semi-analytical formula Eq. (\ref{largeN-b}).

\subsection{Brownian motion with drift}

Now we consider the general case where  the  classical drift is present in addition to the noise term, corresponding to $\dot \phi_0 \neq 0$.   As discussed in Section 3, when the field has the classical velocity 
there  is a limit $\phi = \phi_{\mathrm {max}}$  defined in Eq. (\ref{phi-max}) beyond  which the field cannot go classically.  As argued in \cite{Firouzjahi:2018vet} we can imagine that when the field has approached the classical limit $\phi_{\mathrm {max}}$, then its classical evolution  becomes more and more negligible  and the effects from the diffusion terms  ${\bf W}(\cN)$ becomes more relevant. 

In terms of  $\phi_{\mathrm {max}}$,  Eq. (\ref{25b})  can be cast into
\ba
\label{chi-eq}
\frac{H}{2 \pi \sqrt{c_s}  \phi_{\mathrm {max}}} {\bf W}(\cN) = \chi (\cN) + 
e^{-3 c_s^2 \cN} \, ,
\ea
in which the field displacement relative to $\phi_{\mathrm {max}}$ is defined via
\ba
\chi \equiv \frac{\phi(\cN)}{\phi_{\mathrm {max}}} -1 \, .
\ea

Though  Eq. (\ref{chi-eq}) has a simple form but it can not be solved analytically to obtain 
$p_\pm$ and  $\langle \cN \rangle$. The main reasons are that there is  a time-dependent drift term   $ e^{-3{{c_s^2}} \cN}$  and also  that the stochastic variable $\cN$ appears in ${\bf W}(\cN)$.  This is unlike the case in previous section  where there was  effectively a single barrier, i.e., the surface of end of inflation,  and one could find analytic results. 

As observed in \cite{Firouzjahi:2018vet} in the limit that  the contribution of the drift term $ e^{-3{{c_s^2}} \cN}$ becomes negligible, then Eq. (\ref{chi-eq}) becomes a pure Brownian motion like Eq. (\ref{25c}) 
with the initial condition  $\chi=0$.  Intuitively speaking, this corresponds to the situation when one
starts from $\phi_{\mathrm {max}}$ with zero initial velocity and the evolution of the field is governed by 
 the quantum kicks as in Brownian motion. 
 In order for this approximation to be valid, we have to wait long enough, $\cN \gg1$,  so the classical velocity of the field falls off.  In this limit,   Eq. (\ref{Brownian-probs}) with the replacement $\phi_\pm \rightarrow \chi_\pm$ can be used to obtain $p_\pm$,  yielding 
\ba
\label{p+-app}
p_+ &\simeq& \frac{-\chi_-}{\chi_+ - \chi_-}  = \frac{\dot \phi_0 - 3 H  c_s^2\phi_-}{ 3 H c_s^2 ( \phi_+ - \phi_-)}
\, , \\
\label{p--app}
p_- &\simeq& \frac{\chi_+}{\chi_+ - \chi_-}  = \frac{-\dot \phi_0 + 3 H c_s^2 \phi_+}{ 3 H c_s^2  ( \phi_+ - \phi_-)} \, .
\ea

The analysis of $\langle \cN \rangle$ is more non-trivial due to non-linear effects in the factor $I(\cN)$ as we observed in the case of pure Brownian motion in previous subsection. However, as before,  we obtain the following implicit equation 
\ba
\label{N-av-drift}
\big \langle I(\cN)  \big\rangle  \simeq   \frac{-\chi_- \chi_+}{\big(  \frac{H}{2\pi \sqrt{c_s} \phi_{\mathrm {max}}  } \big)^2}   =  \frac{-\chi_- \chi_+}{ 9 c_s^4 P_{,X} \calP_\calR} \, ,
\ea
where to obtain the final equation the relation $\epsilon = X P_{,X}/H^2$ has been used.  For $\cN \gg 1$ we obtain the following order of magnitude estimation
\ba
\label{largeN-drift}
\langle \cN \rangle \sim \frac{1}{3 (1- c_s^2) } \ln \Big[   \frac{-\chi_- \chi_+}{3 c_s^4  P_{,X} \calP_\calR}   (1- c_s^2) \big)  
\Big] \,     \quad \quad    \big( \cN \gg 1 \big) \, .
\ea

\begin{figure}[t!]
		\centering
		\includegraphics[scale=0.3]{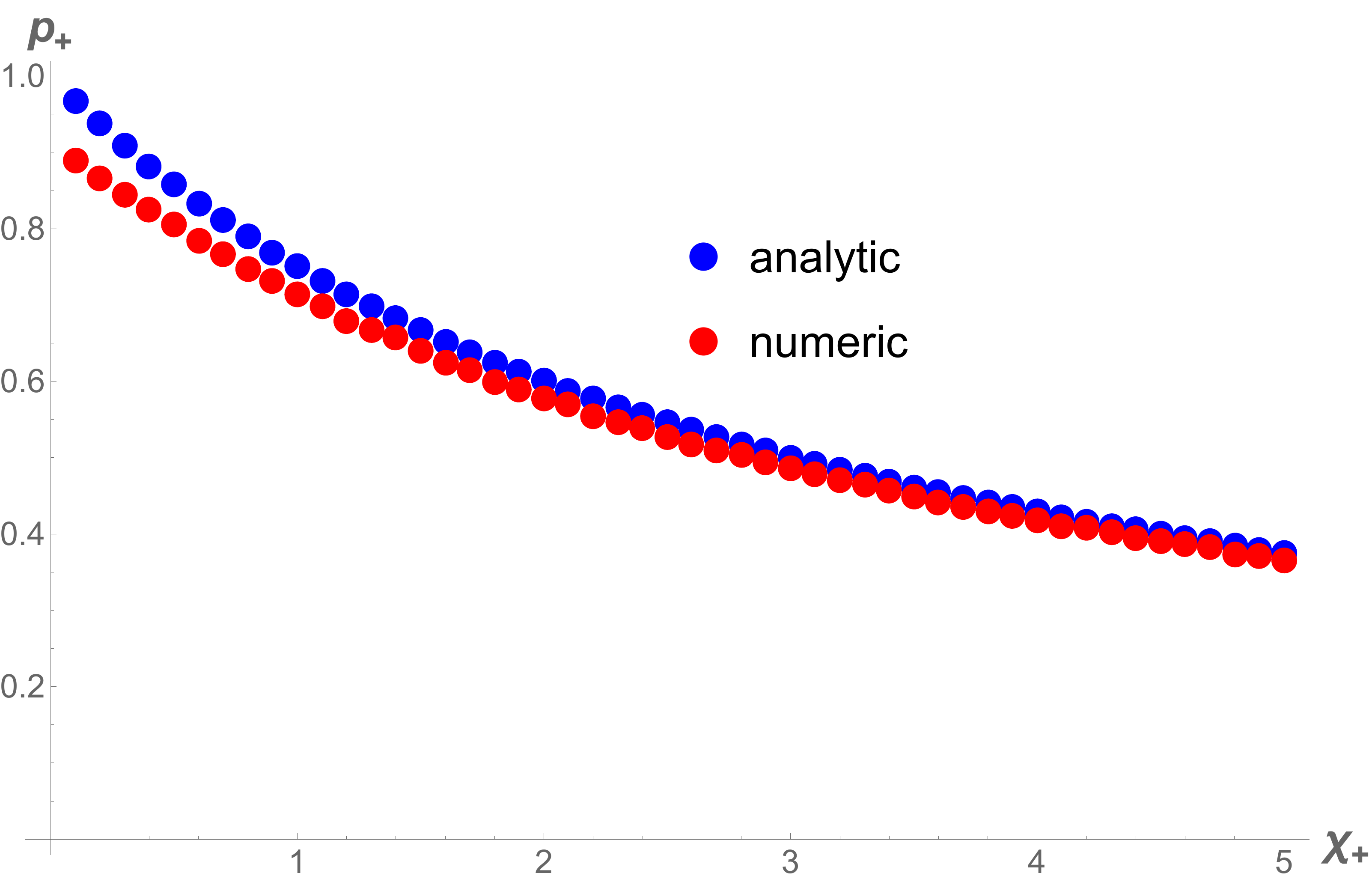}
		\hspace{0cm}
		\includegraphics[scale=0.3]{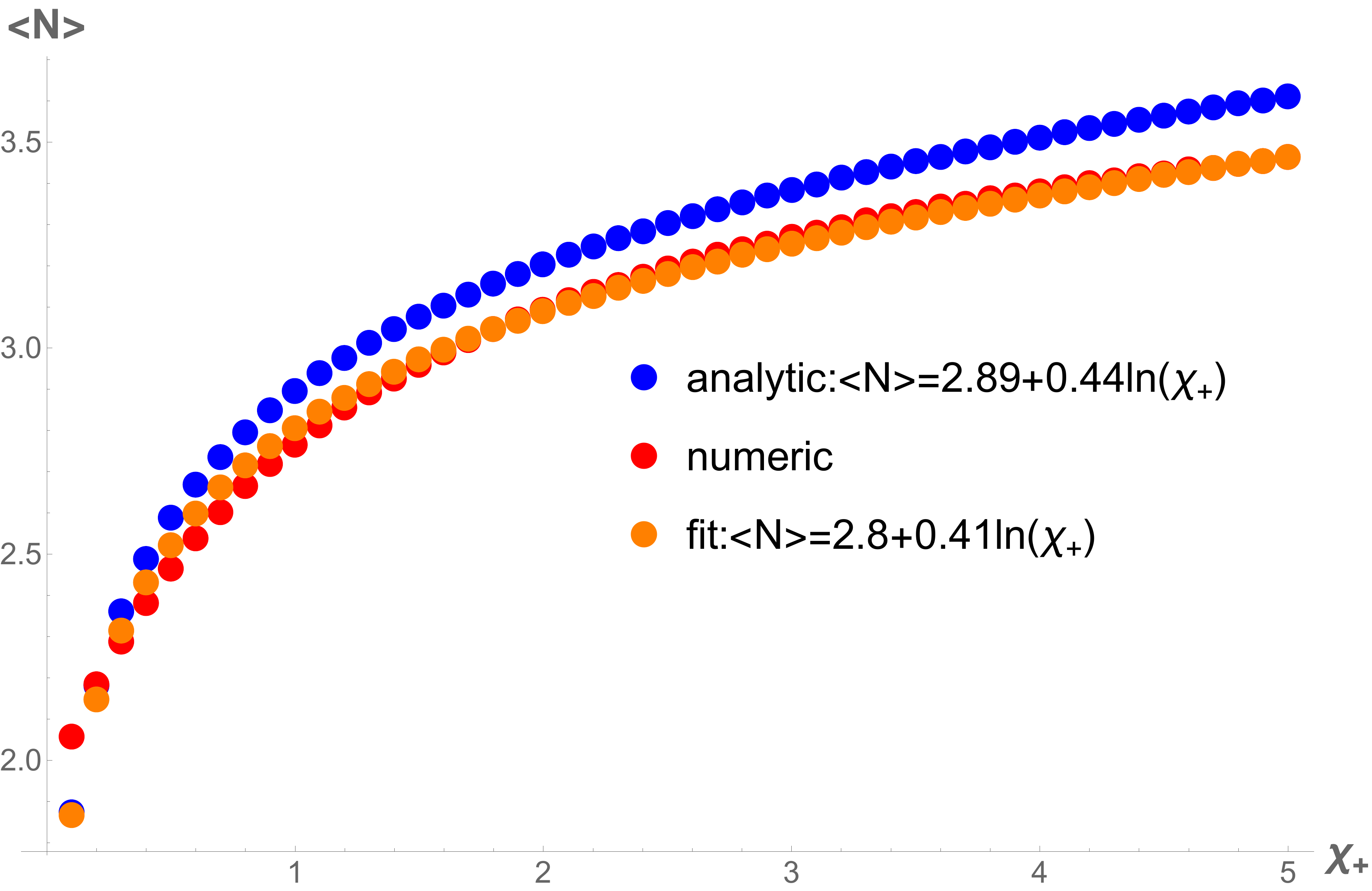}	
		\vspace{1cm}
	\caption{Top: the figure shows the simulated probability  for the case where the drift is not zero. As we see there is good agreement between our analytical approximation  Eq. \eqref{p+-app} and the numeric solution as $\chi_+$ becomes large. Here we have set $\chi_-=-3$ and have varied $\chi_+$ from 0.1 to 5 with $c_s=0.5$ and $\frac{H}{2\sqrt{\pi c_s}}=10^{-1}$. Bottom: as we see the mean number of e-folds estimated by Eq. \eqref{largeN-drift} and the numerical values 
qualitatively look similar  as both show  logarithmic behaviour for large values of $\chi_+$. Moreover a curve is fitted to the numerical data with the numerical coefficients  close to those of Eq.  \eqref{largeN-drift}. }
		\label{probabilitydrift}
\end{figure}

Another  limit that we can calculate $\langle \cN \rangle$ analytically is when the field is released at a distance very close to the right boundary, i.e. 
when $\phi_0 \sim \phi_+$. In our convention, this corresponds to the initial 
condition $\chi\simeq-1$. In this configuration, we expect that the field hits the right boundary in short interval of time, either by classical motion or via one stochastic jump. Taking $\cN\ll1$, from Eq. (\ref{chi-eq}) we find 
\begin{equation}\label{smallNdrift}
    \left<\mathcal{N}\right>\simeq\frac{1+\chi_+}{3c_s^2} \,   \quad \quad    \big( \cN \ll 1 \big) \, .
\end{equation}

In Fig.  \ref{probabilitydrift} we have presented  the simulated probability and mean number of e-folds for the case of Brownian motion with the drift.  Compared to the results of \cite{Firouzjahi:2018vet} we see significant deviation for the behaviour of $\langle \cN \rangle$ for large values of $\chi_+$ for a fixed value of  $\chi_-$.  This is because of the non-linear effects in the parameter $I( \cN)$. In \cite{Firouzjahi:2018vet}, $\langle \cN \rangle$ was evolving linearly in $\chi_+$ as given by Eq. (\ref{smallNdrift}). However, for large $\chi_+$, i.e. for large field displacement, the non-linear effects in $P(X)$ and correspondingly in $I( \cN)$ becomes important and the shape of $\langle \cN \rangle$ deviates significantly from the linear curve, approaching a 
$\log$ curve as our formula Eq. (\ref{largeN-drift}) suggests.

\section{Large Diffusion}
\label{large kappa}

In this section we calculate the power spectrum  in  a setup in which the diffusion (the coefficient of the white noise) is much larger than the drift (the non-stochastic term) in the Langevin equation.  As such, this setup is more akin to eternal inflation where the quantum kicks of the scalar fields dominate over its classical slow-rolling dynamics.

Let us write the solution~\eqref{25} of the Langevin equation (in the case of time-independent diffusion) as follows
\begin{equation}\label{phi-mu-h}
    \phi(N)-\phi_0=\mu F(N)+h {\bf W}(N),
\end{equation}
where $\phi_0\in [\phi_-,\phi_+]$ is the initial condition for the motion and
\begin{equation}
\mu \equiv \frac{\dot\phi_0}{3Hc_s^2}, \qquad
h \equiv \frac{H}{2\pi \sqrt{c_s}}, \qquad
F(N) \equiv 1-\exp\left(-3Nc_s^2\right).
\end{equation}
Our assumption that the diffusion coefficient dominates over the drift coefficient is translated into  $|h| \gg |\mu|$.\footnote{In the previous sections we used the parameter $\kappa$ defined in Eq.~\eqref{kappa-def} as a measure of the diffusion-to-drift ratio (at the boundary crossing time $N=\cal N$).  However, $\kappa$ is defined in terms of $N_c$ which is meaningful only when $\phi_e$ (which corresponds to $\phi_+$) is smaller than $\phi_{\rm max}$, whereas for $\mu=0$ we have $\phi_{\rm max} = \phi_0$, so the interval $\phi_+ \in [\phi_0, \phi_{\rm max}]$ shrinks to zero size.  One can write $\kappa = H / 2\pi\sqrt{c_s} (\phi_{\rm max}-\phi_e)$ without reference to $N_c$, but then it becomes negative when the right boundary $\phi_e=\phi_+$ falls beyond the classically reachable value $\phi_{\rm max}$; then even the magnitude $|\kappa|$ does not provide a good measure of diffusion-to-drift.}.  

We have already considered the boundary crossing probability in this setup in subsection~\ref{ssec:BrownianMotion} so our main objective here is to
calculate  the power spectrum.  In fact, we shall set $\mu=0$ in this section and consider the $O(\mu)$ corrections in appendix~\ref{app:moment}.  This is because the large-diffusion limit is already non-trivial in the leading order.  To see this, note that in the small-diffusion case, almost all trajectories would eventually cross the reheating surface $\phi_e$ at some time.  In other words, the probability to escape to infinity was zero (which is a generic feature of single-field models~\cite{Noorbala:2019kdd}), and that was why we didn't have to impose a left boundary $\phi_-$, i.e., the results were independent of $\phi_-$ in the limit $\phi_-\to-\infty$.  Although Eq.~\eqref{Brownian-probs} shows that, with fixed $\phi_+$, we still have $p_-\to0$ in the limit $\phi_-\to-\infty$, it turns out, as we will see below, that the moments $\langle{\cal N}\rangle$ and $\langle{\cal N}^2\rangle$ diverge so we need to regularize them by imposing both right and left boundaries $\phi_\pm$. In addition, to simplify the analysis further, we set $c_s=1$ corresponding to $P(X)=X$ and 
${\bf W}(N)= W(N)$.  It would be interesting to extend the current analysis to the case with the general value of $c_s$. 

Let us recall Eqs.~\ref{Brownian-probs} and \ref{N-av-Brownian} that we obtained in subsection~\ref{ssec:BrownianMotion}, which in the notation of this section read (note that $I=N$ now):
\begin{equation}\label{61}
    \left<\mathcal{N}\right>=-\frac{(\phi_+-\phi_0)(\phi_--\phi_0)}{h^2},
\end{equation}
and 
\begin{equation}\label{62}
\begin{split}
   p_+ = \frac{-\phi_-+\phi_0}{\phi_+-\phi_-}, \quad \quad  p_- = \frac{\phi_+-\phi_0}{\phi_+-\phi_-}.
    \end{split}
\end{equation}
Note that because of shift symmetry, we can set $\phi_0=0$ as in the analysis in previous sections,  but for the later use we have kept it arbitrary.

We will also need a couple of stochastic identities~\cite{Firouzjahi:2018vet} that we collect here for future reference:
\begin{equation}\label{W3}
    \left<W(\mathcal{N})^3\right>=3\left<\mathcal{N} W(\mathcal{N})\right>,
\end{equation}
\begin{equation}\label{W4}
    \left<W(\mathcal{N})^4\right>=6\left<\mathcal{N} W(\mathcal{N})^2\right>-3\left<\mathcal{N}^2\right>,
\end{equation}
\begin{equation}\label{W5}
    \left<W(\mathcal{N})^5\right>=10\left<\mathcal{N} W(\mathcal{N})^3\right>-15\left<\mathcal{N}^2 W(\mathcal{N})\right>.
\end{equation}

Let us calculate $\left<\mathcal{N}_+\right>$ and $\left<\mathcal{N}_-\right>$ first, which will prove to be handy for further calculations.  $\left<\mathcal{N}_+\right>$ is the average number of $e$-folds to hit $\phi_+$ over all trajectories that are conditioned such that $\phi_+$ is hit earlier than $\phi_-$.  Obviously, we have
\begin{equation}\label{condn}
     \left<\mathcal{N}\right>=p_+\left<\mathcal{N}_+\right>+p_-\left<\mathcal{N}_-\right>.
\end{equation}
Moreover, evaluating Eq.~\eqref{phi-mu-h} at $\cal N$, raising it to the third power and then taking the average, we find
\begin{equation}\label{cubed}
    p_+(\phi_+-\phi_0)^3+p_-(\phi_--\phi_0)^3 = h^3\left<W(\mathcal{N})^3\right>.
\end{equation}
We can simplify the right hand side using Eqs.~\eqref{W3} by first writing
\begin{equation}
    \left<W(\mathcal{N})^3\right> = 3\Big< \mathcal{N} \frac{\phi(\mathcal{N})-\phi_0}{h} \Big> = 3\Big(p_+\frac{\phi_+-\phi_0}{h}\left<\mathcal{N}_+\right>+p_-\frac{\phi_--\phi_0}{h}\left<\mathcal{N}_-\right>\Big),
\end{equation}
and then using Eq.~\eqref{cubed} to obtain
\begin{equation}\label{N+2}
    p_+(\phi_+-\phi_0)^3+p_-(\phi_--\phi_0)^3=
    3h^2 \Big( p_+\left<\mathcal{N}_+\right>(\phi_+-\phi_0)+p_-\left<\mathcal{N}_-\right>(\phi_--\phi_0) \Big).
\end{equation}
Now Eqs.~\eqref{condn} abd \eqref{N+2} form a system of two equations for $\left<\mathcal{N}_\pm\right>$ which can readily be solved to yield
\begin{equation}\label{N+C}
    \left<\mathcal{N}_+\right>=\frac{\left(\phi _+-\phi _0\right) \left(\phi _+-2 \phi _-+\phi _0\right)}{3 h ^2},
\end{equation}
\begin{equation}\label{N-C}
    \left<\mathcal{N}_-\right>=\frac{\left(\phi _--\phi _0\right) \left(\phi _--2 \phi _++\phi _0\right)}{3 h ^2}.
\end{equation} 

In order to compute the power spectrum, we need to calculate $\left<\mathcal{N}^2\right>$. To this end, we start by evaluating Eq.~\eqref{phi-mu-h} at $\cal N$, raising it to the fourth power and then taking the average:
\begin{equation}\label{quartic}
    p_+(\phi_+-\phi_0)^4+p_-(\phi_--\phi_0)^4=h^4\left<W(\mathcal{N})^4\right>.
\end{equation}
Using Eq.~\eqref{W4}, as well as
\begin{equation}\label{nw2}
    \big<\mathcal{N} W(\mathcal{N})^2\big> = \Big<\mathcal{N}\frac{(\phi(\mathcal{N})-\phi_0)^2}{h^2}\Big> = p_+\big<\mathcal{N}_+\big>\frac{(\phi_+-\phi_0)^2}{h^2} + p_-\big<\mathcal{N}_-\big>\frac{(\phi_--\phi_0)^2}{h^2},
\end{equation}
we can read $\langle {\cal N}^2 \rangle$ from Eq.~\eqref{quartic} as
\begin{equation}\label{n2}
    \left<\mathcal{N}^2\right>=  \frac{-1}{3 h^4} \Big[ 
    {(\phi _--\phi_0)( \phi _+-\phi_0) \left[ (\phi _--\phi_0)^2-3 (\phi _+-\phi_0)(\phi _--\phi_0)+(\phi _+-\phi_0)^2 \right]} \Big] \, .
\end{equation}

We will also be interested in calculating $\left<\mathcal{N}_\pm^2\right>$. Evidently, they are related to $\left<\mathcal{N}^2\right>$ via
\begin{equation}\label{condn2}
    \left<\mathcal{N}^2\right>=p_+\left<\mathcal{N}^2_+\right>+p_-\left<\mathcal{N}^2_-\right>.
\end{equation}
To proceed, we evaluate Eq.~\eqref{phi-mu-h} at $\cal N$, raise it to the fifth power and then take the average:
\begin{equation}\label{fifth}
    p_+(\phi_+-\phi_0)^5+p_-(\phi_--\phi_0)^5=h^5\left<W(\mathcal{N})^5\right>.
\end{equation}
To calculate the right hand side using Eq.~\eqref{W5}, we need
\begin{equation}\label{nw3}
    \big<\mathcal{N}W(\mathcal{N})^3\big>=\big<\mathcal{N}\frac{(\phi(\mathcal{N})-\phi_0)^3}{h^3}\big>=p_+\big<\mathcal{N}_+\big>\frac{(\phi_+-\phi_0)^3}{h^3}+p_-\left<\mathcal{N}_-\right>\frac{(\phi_--\phi_0)^3}{h^3},
\end{equation}
and
\begin{equation}\label{wn2}
    \big<\mathcal{N}^2 W(\mathcal{N})\big>=\big<\mathcal{N}^2\frac{(\phi(\mathcal{N})-\phi_0)}{h}\big>=p_+\big<\mathcal{N}^2_+\big>\frac{(\phi_+-\phi_0)}{h}+p_-\big<\mathcal{N}^2_-\big>\frac{(\phi_--\phi_0)}{h}.
\end{equation}
Plugging Eqs.~\eqref{nw3} and \eqref{wn2} in Eq.~\eqref{W5} and then in Eq.~\eqref{fifth}, we get an equation for $\left<\mathcal{N}^2_\pm\right>$.  Together with Eq.~\eqref{condn2}, it provides a system of equations which can be readily solved for $\left<\mathcal{N}^2_\pm\right>$ to yield
\begin{equation}
{  \left<\mathcal{N}_+^2\right>= \frac{-1}{45 h^4} \Big[ 
   {\left(2 \phi _--\phi _+-\phi _0\right) \left(\phi _+-\phi _0\right) \left(4 \phi _-^2+\left(6 \phi _0-14 \phi _+\right) \phi _-+7 \phi _+^2-3 \phi _0^2\right)} \Big] \, ,}
\end{equation}
and 
\begin{equation}
 { \left<\mathcal{N}_-^2\right>= \frac{-1}{45 h^4}
   \Big[ {\left(2 \phi _+-\phi_ --\phi _0\right) \left(\phi _--\phi _0\right) \left(4 \phi _+^2+\left(6 \phi _0-14 \phi _-\right) \phi _++7 \phi _-^2-3 \phi _0^2\right)} \Big]  \, .}
\end{equation}
As we will see in appendix~\ref{app:moment} one can get the same results by using the distribution function directly.

Finally we can calculate the power spectrum, which in the stochastic $\delta\mathcal{N}$ formalism is given by
\begin{equation}
  \label{power-deltaN}
\mathcal{P_\calR}=\frac{d\left<\delta\mathcal{N}^2\right>}{d\left<\mathcal{N}\right>}=\frac{d\left<\delta\mathcal{N}^2\right> / d\phi_0}{d\left<\mathcal{N}\right> / d\phi_0},
\end{equation}
where we remind that $\langle\delta\mathcal{N}^2\rangle = \langle  \cN^2 \rangle - \langle \cN \rangle^2 $.  It is therefore enough to insert the formulas for $\left<\mathcal{N}\right>$ and $\left<\mathcal{N}^2\right>$ that we have obtained above to get
\begin{equation}
\label{power-N}
\mathcal{P_\calR}=\frac{\left(2 \phi _0 - \phi _- - \phi _+ \right)^2}{3 h^2}.
\end{equation}
There is a subtlety here. Since $\mathcal{N}$ is defined as the number of $e$-folds to either $\phi_+$ or $\phi_-$, the resulting power spectrum of curvature perturbations (which are created during inflation on a flat potential) is calculated on the surface $\Sigma = \{ \phi=\phi_+ \} \cup \{ \phi=\phi_- \}$.  This can be the case when reheating takes place on $\Sigma$, i.e., on both sides $\phi_\pm$ of the starting point $\phi_0$.  Another case of interest to which Eq.~\eqref{power-N} applies is when the phase of ultra-slow-roll terminates on both $\phi_\pm$.  

However, if the end of inflation (or end of the ultra-slow-roll phase) occurs only on $\phi_+$, i.e., if $\Sigma = \{ \phi=\phi_+ \}$ then we need to consider only those trajectories that reach this new $\Sigma$.  In this situation, we must replace $\left<\mathcal{N}\right>$ in Eq.~\eqref{power-N} with $\left<\mathcal{N_+}\right>$ ($\phi_-$ will then play the role of a cut-off) and so we get
\begin{equation}\label{power-N+}
\mathcal{P_\calR}=\frac{4 \left(\phi _0 - \phi _-\right)^2}{15 h^2}.
\end{equation}

Let us also compare these results with that of Ref.~\cite{Pattison:2017mbe} where a reflective boundary condition is used on $\phi_-$ and the power spectrum is calculated to be\footnote{To convert ${\cal P}_\zeta = 2 \mu^2 (x-1)^2/3$ (Eq.~(5.8) of Ref.~\cite{Pattison:2017mbe} to our notation here, use their definitions $\mu^2=(\phi_+-\phi_-)^2/vM_P^2$ and $x=(\phi_0-\phi_+)/(\phi_+-\phi_-)$, yielding Eq.~\eqref{power-reflective}.}
\begin{equation}\label{power-reflective}
{\cal P}_\calR = \frac{4 \left(\phi _0 - 2\phi _+ + \phi_- \right)^2}{3 h^2}.
\end{equation}
In all three cases, Eqs.~\eqref{power-N}, \eqref{power-N+}, and \eqref{power-reflective} the power spectrum is proportional to $h^{-2} \propto H^{-2}$, and for large values of $\phi_+ - \phi_0$ or $ \phi_0 -\phi_-$, to $(\phi_+ - \phi_0)^2 $ or $ (\phi_- - \phi_0)^2$.  However, the numerical prefactors differ, as well as the exact dependence on the boundary values $\phi_\pm$.  This is not surprising, as these three cases correspond to three physically different situations.  As we mentioned before, Eq.~\eqref{power-N} corresponds to $\Sigma$ (the end of inflation, or end of ultra-slow-roll phase) being when $\phi$ reaches either $\phi_+$ or $\phi_-$.  In the approach of Ref.~\cite{Pattison:2017mbe}, this corresponds to absorbing boundary conditions on both $\phi_\pm$.  On the other hand, Eq.~\eqref{power-N+} corresponds to a multiverse setup that exiting through $\phi_+$ and $\phi_-$ leads to two different universes and we want to calculate the power spectrum in the universe where the right boundary $\phi_+$ has been hit.  Finally, Eq.~\eqref{power-reflective} applies when there is a reflecting wall (e.g., a very high potential barrier) at $\phi_-$ that ensures that all trajectories exit inflation through $\phi_+$.

\section{Summary and Discussions}
\label{summary}

In this paper we have extended the stochastic inflation formalism to the setup of non-attractor inflation. To simplify the analysis we have restricted ourselves to the $P(X)$ setup which is shift symmetric. We have obtained the Langevin equations for the long mode perturbations and have solved them. In principle these analysis can be extended to a more general $P(X, \phi)$ setup though the analysis will be more 
complicated.  

We have calculated the stochastic corrections in cosmological observables such as 
$\langle \cN \rangle$ and the curvature perturbation power spectrum $\calP_\calR$. We have shown that the stochastic corrections in these observables are sub-leading. More specifically, it is shown that the fractional corrections in these observables are at the order of power spectrum. Compared to analysis in USR setup, the non-linearities  inherited in $P(X)$ setup and the effects of $c_s$ made the analysis of  stochastic correlations such as  $\langle {\bf W (\cN)}^n \rangle$ more non-trivial and we had to employ the Ito lemma iteratively to calculate various stochastic integrals. 

We also studied the boundary crossings and the first hitting probabilities in a hypothetical setup in dS space with two boundaries in field space.  We have calculated the probabilities of hitting first either the right or left boundaries in a pure Brownian case where the field has no classical drift and moves only subject to stochastic kicks. The general case when both the classical drift and the stochastic diffusion terms are present is more non-trivial. However, in the large $\cN$ limit this configuration approaches a Brownian limit where the classical motion reaches its final limit in field space and the subsequent dynamics is controlled by the diffusion term. We have calculated the first hitting probabilities in this limit as well and have verified that the results are in good agreements with the full numerical simulations.

Finally, we considered a hypothetical setup in which the diffusion term dominates over the classical drift term. This situation is more akin to the eternal inflation picture where the  quantum kicks push the field upwards and prevent it from the classical slow-rolling. Of course, this setup is opposite to the conventional regime of inflation where the classical slow-rolling is the dominant effects in the dynamics of the field. 
We have
calculated the power spectrum and have verified that it scales like
$H^{-2}$, although its exact form depends on the details of the
boundary conditions imposed. This can have interesting implications for eternal inflation and for the collapse of regions of spacetime to form primordial black hole during eternal inflation.

\vspace{0.5cm}

 {\bf Acknowledgments:}  We are grateful to Hooshyar Assadullahi, Vincent Vennin and David Wands  for discussions.  A. N. would like to thank APC, Paris 7 and  Tokyo Institute of Technology (TITECH) for hospitalities while this work was in progress.  M.N. acknowledges financial support from the research council of University of Tehran.

\appendix
\section{Stochastic Integrals}
\label{calculus}

In this appendix we provide the details of the stochastic analysis to calculate various correlation functions perturbatively in powers of $\kappa$. 

Our starting point is the perturbative expansion of $\cN$ in powers of $\kappa$, 
\begin{equation}
\label{27b}
    \mathcal{N}=N_c+\frac{1}{3c_s^2}\sum^{\infty}_{n=0}\frac{(-\kappa)^n}{n}  {\bf W}(\cN)^n \, ,
\end{equation}
in which
\ba
\label{hatW-b}
{\bf W}(\cN) \equiv \int_0^\cN \frac{d W(N)}{\sqrt{P_{,X}(N)}} \, , \quad \quad 
d W(N) \equiv \xi(N) dN \, .
\ea
As mentioned in the main text, $W(\cN)$ is the Wiener process \cite{Evans} associated with the noise $\xi(N)$ satisfying 
\ba
\label{Weiner}
\langle W(\cN) \rangle =0 \, , \quad \quad  \langle W(\cN)^2 \rangle = \langle \cN \rangle \, .
\ea
To calculate power spectrum, we need to calculate $\langle \cN \rangle$ and $\delta \cN^2=  \langle \cN^2  \rangle - \langle \cN \rangle^2$ to order $\kappa^4$ in which 
\ba
\label{N-av-b}
\langle \cN \rangle = N_c + \frac{\kappa^2}{6 c_s^2} \big \langle {{\bf W}(\cN)}^2  \big \rangle
-\frac{\kappa^3}{9 c_s^2} \big \langle {{\bf W}(\cN)}^3  \big\rangle  +  \frac{\kappa^4}{12 c_s^2} \big \langle {{\bf W}(\cN)}^4  \big \rangle +O(\kappa^5)
\ea
and
\ba
\label{deltaN2-b}
\delta \cN^2 =  \frac{\kappa^2}{9 c_s^4} \big \langle {{\bf W}(\cN)}^2 \big \rangle
-\frac{\kappa^3}{9 c_s^4} \big \langle {{\bf W}(\cN)}^3 \big \rangle  +  \frac{\kappa^4}{108 c_s^4} \Big[ 11\langle {{\bf W}(\cN)}^4 \big \rangle - 3\big \langle {{\bf W}(\cN)}^2 \big \rangle ^2\Big]  +O(\kappa^5)
\ea
Consequently, we have to calculate $\langle {{\bf W}(\cN)}^2 \rangle, \langle {{\bf W}(\cN)}^3 \rangle$ and
$\langle {{\bf W}(\cN)}^4 \rangle$ to orders of $\kappa^2, \kappa$ and $\kappa^0$ respectively. 
For this purpose, we use the following 
fundamental properties of the stochastic integral  \cite{Evans}
\begin{equation}
\label{29b}
\Big\langle \int^\mathcal{\mathcal{N}}_0f\left(N\right)dW\Big\rangle=0 \, ,
\end{equation}
and the Ito lemma, in which states that \cite{Evans, Firouzjahi:2018vet}
\ba
\label{Ito}
d \big ( W(N)^n \big) = n W{{(N)}}^{n-1} d W + \frac{n (n-1)}{2} W{{(N)}}^{n-2} d N \, .
\ea
Now taking the stochastic average of the integrated form of  the above formula and using Eq. (\ref{29b}) yields \begin{equation}
\label{45b}
     \Big \langle {{{{\bf W}(\cN)}^n}} \Big\rangle=\frac{n(n-1)}{2} \Big \langle \int^\mathcal{N}_0 \frac{{\bf W}\left(N\right)^{n-2}}{P_{,X}}dN \Big\rangle \, .
\end{equation}
The advantage of the above formula is that it relates $ \big \langle {{\bf W}(\cN)}^n \big \rangle$ to $\big \langle {{\bf W}(\cN)}^{n-2} \big \rangle$. In particular, for $n=2, 3$ and $4$ we obtain 
\ba
\label{W2}
\big \langle {{\bf W}(\cN)}^2 \big \rangle = \Big \langle     \int_0^{\cN} \frac{d N}{P_{,X}(N)}
\Big \rangle 
\ea
\begin{equation}
\label{45p}
     \big \langle{{{{\bf W}(\cN)}^3}}\big\rangle=3\Big \langle \int^\mathcal{N}_0 \frac{{\bf W}\left(N\right)}{P_{,X}}  dN \Big\rangle 
\end{equation}
\begin{equation}
\label{46}
   \big \langle{{{{\bf W}(\cN)}^4}} \big\rangle=6\Big \langle \int^\mathcal{N}_0 \frac{{\bf W}\left(N\right)^2}{P_{,X}}  dN \Big\rangle 
\end{equation}
Note that $\cN$ is a stochastic variable, so the above stochastic integrals are non-trivial. 

We proceed iteratively in powers of $\kappa$. First, from Eq. (\ref{N-av-b}) we have 
\ba
\big \langle \cN \big \rangle = N_c + \frac{\kappa^2}{6 c_s^2} \big \langle {{\bf W}(\cN)}^2 \big \rangle + O(\kappa^4)\, .
\ea
Combining with Eq. (\ref{W2}), to order $\kappa^2$ we obtain 
\ba
\label{N-av-c}
\big \langle \cN \big \rangle = N_c + \frac{\kappa^2}{9 c_s^4}  I(N_c)+ O(\kappa^4)\, ,
\ea
where to simplify the notation, we have defined the integral $I(N)$ as
\ba
I(N) \equiv \int_0^{ N } \frac{dN}{P_{,X}(N)} \, .
\ea
Note the important step in which the upper limit of the integral in Eq. (\ref{N-av-c}) is replaced by $N_c$ which is valid to  order $\kappa^2$. 

Similarly,  to leading order for $\delta \cN$ we have 
\ba
\delta \cN = \cN - \langle \cN \rangle = -\frac {\kappa}{3 c_s^2} {{\bf W}(\cN)} + O(\kappa^2) \, ,
\ea
yielding
\ba
\big \langle \delta \cN^2  \big \rangle =  \frac{\kappa^2}{9 c_s^4} \Big\langle  \int_0^\cN \frac{dN}{P_{,X}}  \Big\rangle + O(\kappa^4) =  \frac{\kappa^2}{9 c_s^4} I(N_c)  + O(\kappa^4) \, .
\ea
Now we calculate $\big \langle {{\bf W}(\cN)}^2 \big \rangle$ to order $\kappa^2$. From Eq. (\ref{W2}) we have
\ba
\label{pert1}
\big \langle {{\bf W}(\cN)}^2  \big \rangle = \Big \langle     \int_0^{\cN} \frac{d N}{P_{,X}(N)}
\Big \rangle =  \Big \langle     \int_0^{\langle \cN \rangle} \frac{d N}{P_{,X}(N)}
\Big \rangle +  
 \Big \langle     \int_{\langle \cN \rangle}^{\langle \cN \rangle + \delta \cN} \frac{d N}{P_{,X}(N)}
\Big \rangle \, .
\ea
Using the perturbative expansion Eq. (\ref{N-av-c}) for $\langle \cN \rangle$,  the first integral above is calculated to be 
\ba
\Big \langle     \int_0^{\langle \cN \rangle} \frac{d N}{P_{,X}(N)} \Big \rangle 
= \Big (  1+ \frac{\kappa^2}{6 c_s^2 P_{,X}}   \Big) I(N_c)  \, .
\ea
As for the second integral in Eq. (\ref{pert1}) we change the integration variable as follows
\ba
 \Big \langle     \int_{\langle \cN \rangle}^{\langle \cN \rangle + \delta \cN} \frac{d N}{P_{,X}(N)}
\Big \rangle &=&  \Big \langle     \int_{0}^{ \delta \cN} \frac{d T }{P_{,X}( \langle \cN \rangle + T)}
\Big \rangle \nonumber\\
&\simeq&   \Big \langle     \int_{0}^{ \delta \cN} \frac{d T }{P_{,X}( \langle \cN \rangle)} 
\Big(  1- T \frac{P_{,X}'}{P_{,X}} \Big) \Big \rangle \nonumber\\
&\simeq& \frac{P_{,X}' \big( \langle \cN \rangle \big)}{2 P_{,X} \big( \langle \cN \rangle \big)^2} \langle \delta \cN^2 \rangle  \, ,
\ea
in which a prime indicates the derivative with respect to $N$. Note that the above procedure is valid to order $\kappa^2$. 

Combining the above results, to order $\kappa^2$ we obtain
\ba
\big \langle {{\bf W}(\cN)}^2 \big \rangle = \Big( 1+ \frac{\kappa^2}{6 P_{,X} c_s^2} - \frac{\kappa^2 P_{,X}'}{18 c_s^4P_{,X}^2} \Big) I(N_c) + O(\kappa^4) \, .
\ea
Now, using $P_{,X}'= X' P_{XX} = - 6 c_s^2 X P_{XX}$ and the formula for $c_s$ in Eq. (\ref{cs}), the above expression simplifies to
\ba
\big \langle {{\bf W}(\cN)}^2 \big \rangle = \Big( 1+ \frac{\kappa^2}{6 P_{,X} c_s^4}  \Big) I(N_c) + O(\kappa^4) \, .
\ea

To calculate $\big \langle {{\bf W}(\cN)}^4 \big \rangle$, we start from Eq. (\ref{46}) and use the Ito lemma again.
Defining $dQ \equiv dN/P_{,X}$, then the right hand side of Eq. (\ref{46}) is manipulated to
\ba
\big \langle \int_0^\cN {{\bf W}(N)}^2 d Q  \big\rangle  &=&  \Big \langle \Big [ \int_0^{\cN}  d \big({{\bf W}(N)}^2 Q \big) - Q \,  d\big( {{\bf W}(N)}^2 \big)    \Big] \Big \rangle   \nonumber\\
&=&    \Big \langle {{\bf W}(\cN)}^2 Q(\cN)  \Big \rangle -  \Big \langle   \int_0^{\cN} Q(N)  \Big[  2 {\bf W}(N) d {\bf W}(N) + \frac{dN}{P_{,X}} \Big] \Big \rangle \nonumber\\
&=&   \Big \langle {{\bf W}(\cN)}^2 Q(\cN)  \Big \rangle -  \Big \langle   \int_0^{\cN} Q(N)  \Big[  2 {\bf W}(N) d {\bf W}(N) + \frac{dN}{P_{,X}} \Big] \Big \rangle \nonumber\\
&=&   \Big \langle {{\bf W}(\cN)}^2 Q(\cN)  \Big \rangle -    \Big \langle   \int_0^{\cN} Q(N)  \frac{dN}{P_{,X}}  \Big \rangle \, .
\ea
Since we need to calculate $\big  \langle {{\bf W}(\cN)}^4 \big  \rangle$ to order $\kappa^0$, we can simply replace the upper bound for the integrals above by $N_c$, obtaining
\ba
\big  \langle {{\bf W}(\cN)}^4 \big  \rangle = 6 I(N_c)^2 - 6 \int_0^{N_c} \frac{dN}{P_{,X}(N)} I(N) \, .
\ea

Finally, we need to calculate $\big  \langle {{\bf W}(\cN)}^3 \big  \rangle$ to order $\kappa$. Using the Ito lemma (as in above analysis)  for the right hand side of Eq. (\ref{45p}) we obtain
\ba
\Big \langle \int^\mathcal{N}_0 \frac{{\bf W}\left(N\right)}{P_{,X}}  dN \Big\rangle  = \Big \langle  {\bf W}\left(\cN\right)\int^\mathcal{N}_0 \frac{1}{P_{,X}}  dN \Big\rangle 
\ea
Now, we eliminate ${\bf W}(\cN)$ above in favour of $\delta \cN$ as follows. Consider the expansion of Eq. (\ref{27b}) to order $\kappa$, we have
\ba
3 c_s^2 ( \cN - N_c) = - \kappa {\bf W}(\cN) + O(\kappa^2) \quad   \rightarrow \quad
{\bf W}(\cN) = -\frac{3 c_s^2}{\kappa}  \delta \cN + O(\kappa) \, .
\ea
As a result, from Eq. (\ref{45p}) we obtain 
\ba
\langle {{\bf W}(\cN)}^3 \rangle &=& -\frac{9 c_s^2}{\kappa} \Big \langle \delta \cN \int_0^\cN \frac{dN}{P_{,X}}  \Big \rangle  + O(\kappa^3) \nonumber\\
&=&  -\frac{9 c_s^2}{\kappa} \Big \langle \delta \cN \int_0^{ \langle \cN \rangle + \delta \cN}\frac{dN}{P_{,X}}  \Big \rangle  + O(\kappa^3) \nonumber\\
&=& -\frac{9 c_s^2}{\kappa} \Big \langle \delta \cN \Big(  \int_0^{ \langle \cN \rangle }\frac{dN}{P_{,X}}   
+ \frac{\delta \cN}{P_{,X}(\langle \cN \rangle)}  \Big)  \Big \rangle   + O(\kappa^3) \nonumber\\
&=&  -\frac{9 c_s^2}{\kappa P_{,X}(\langle \cN \rangle) } \big \langle \delta \cN^2  \big\rangle  + O(\kappa^3) \nonumber\\
&=&  -\frac{\kappa}{c_s^2 P_{,X}(\langle \cN \rangle)} I(N_c)  + O(\kappa^3)\, .
\ea

Having calculated $\langle {{\bf W}(\cN)}^2 \rangle, \langle {{\bf W}(\cN)}^3 \rangle$ and
$\langle {{\bf W}(\cN)}^4 \rangle$ to necessary orders in $\kappa$, we obtain $\langle \cN \rangle$ and
$\delta \cN^2$ to order $\kappa^4$ as given in Eqs. (\ref{28}) and (\ref{deltN-2}) respectively. 

\section{ Drift Corrections to Large Diffusion }
\label{app:moment}

In this appendix we  calculate the $O(\mu)$ corrections to the quantities $p_\pm$ and $\left<\mathcal{N}\right>$ of section~\ref{large kappa}. To this end we start by taking the average  value of Eq.~\eqref{phi-mu-h} and its square up to first order of $\mu$:
\begin{equation}\label{63}
    p_+(\phi_+-\phi_0)+p_-(\phi_--\phi_0)=\mu\left<F(\mathcal{N})\right>,
\end{equation}
\begin{equation}\label{64}
    p_+(\phi_+-\phi_0)^2+p_-(\phi_+-\phi_0)^2=2\mu h\left<F(\mathcal{N})W(\mathcal{N})\right>+h^2\left<\mathcal{N}\right>+O(\mu^2).
\end{equation} 
As it can be seen from Eqs.~$\eqref{63}$ and $\eqref{64}$ to calculate the first order of $\mu$ contribution, it is enough to calculate $\left<F(\mathcal{N})\right>$ and $\left<F(\mathcal{N})W(\mathcal{N})\right>$ at the zeroth order, i.e., in the standard Brownian regime with no drift. This is because one expects that in the limit $\mu\rightarrow0$ we get the  results in the standard Brownian motion without drift. So the expectation is that if  one  expands $\left<F(\mathcal{N})\right>(\mu)$ in terms of $\mu$ powers then the zeroth order corresponds to the  case without drift. To calculate these two quantities we use the probability distribution function of pure Brownian motion.
It can be shown that in the standard Brownian motion, the probability distribution function $f(N;\phi_+,\phi_-,\phi_0)$ for first crossing any of the barriers $\phi_\pm$ at time $\mathcal{N}=N$, starting from $\phi_0$, has the following form\cite{Karatzas}:
\begin{equation}
\begin{split}\label{67}
  &\sqrt{2\pi h^2N^3}f(N;\phi_+,\phi_-,\phi_0)=\\ &\sum_{n=-\infty}^{\infty}(2n(\phi_+-\phi_-)+(\phi_0-\phi_-))\exp \Big(-\frac{(2n(\phi_+-\phi_-)+(\phi_0-\phi_-))^2}{2h^2N} \Big)\\+&\sum_{n=-\infty}^{\infty}(2n(\phi_+-\phi_-)+(\phi_+-\phi_0))\exp\Big(-\frac{(2n(\phi_+-\phi_-)+(\phi_+-\phi_0))^2}{2h^2N}\Big).
   \end{split}
\end{equation}

Now let's define the following distribution functions
\begin{equation}
\begin{split}
&f_+(N;\phi_+,\phi_-,\phi_0)dN=P[\mathcal{N}_{\phi_+}\in (N,N+dN),\mathcal{N}_{\phi_-}>\mathcal{N}_{\phi_+}],\\
&f_-(N;\phi_+,\phi_-,\phi_0)dN=P[\mathcal{N}_{\phi_-}\in (N,N+dN),\mathcal{N}_{\phi_+}>\mathcal{N}_{\phi_-}],
    \end{split}
\end{equation}
where $\mathcal{N}_{\phi_{\pm}}$ are defined as follows
\begin{equation}\label{65}
\begin{split}
    \mathcal{N}_{\phi_+}=\text{inf}\left\{N\geq 0|\phi(N)>\phi_+\right\},\\
    \mathcal{N}_{\phi_-}=\text{inf}\left\{N\geq 0|\phi(N)<\phi_-\right\}.
    \end{split}
\end{equation}
In other words $f_\pm$ are the time distribution functions such that one of the barriers is hit earlier than the other one. One can verify that 
\begin{equation}
    f(N;\phi_+,\phi_-,\phi_0)=f_+(N;\phi_+,\phi_-,\phi_0)+f_-(N;\phi_+,\phi_-,\phi_0).
\end{equation}
Moreover by \eqref{65} it is clear that
\begin{equation}
    \mathcal{N}=\text{min}(\mathcal{N}_{\phi_+},\mathcal{N}_{\phi_-}).
\end{equation}
It can be shown that 
\begin{equation}\label{f-}
\begin{split}
  &f_-(N;\phi_+,\phi_-,\phi_0)=\\&\frac{1}{\sqrt{2\pi h^2N^3}} \sum_{n=-\infty}^{\infty}(2n(\phi_+-\phi_-)+(\phi_0-\phi_-))\exp \Big(-\frac{(2n(\phi_+-\phi_-)+(\phi_0-\phi_-))^2}{2h^2N} \Big),
  \end{split}
\end{equation}
\begin{equation}\label{f+}
\begin{split}
  &f_+(N;\phi_+,\phi_-,\phi_0)=\\&\frac{1}{\sqrt{2\pi h^2N^3}} \sum_{n=-\infty}^{\infty}(2n(\phi_+-\phi_-)+(\phi_+-\phi_0))\exp \Big(-\frac{(2n(\phi_+-\phi_-)+(\phi_+-\phi_0))^2}{2h^2N} \Big).
  \end{split}
\end{equation}
In passing, note that unlike $f$ that is normalized to one, $f_\pm$ are normalized to $p_\pm$.

Before evaluating the needed expectations it is useful to introduce the generating function $M$ for the moments of an arbitrary random variable like $X$ as follows: 
\begin{equation}
\label{moment}
   M(s)=\left<\exp(sX)\right>=\int_X\exp(sx) f_X(x)dx,
\end{equation}
where the integral is calculated over the domain of $X$ and $s\leq0$. Also we have
\begin{equation}
    \lim_{s\rightarrow0}M'(s)=\lim_{s\rightarrow0}\int_X x\exp(sx) f_X(x)dx=\left<X\right>.
\end{equation}
It is straightforward to generalize the above formula and write
\begin{equation}
   \lim_{s\rightarrow0} M^{(n)}(s)=\left<X^n\right>,
\end{equation}
where by $M^{(n)}(s)$ we mean the $n$-th derivative of $M(s)$. So it is easy to calculate $M(s)$ once and then we can get the $n$-th moment of $X$, i.e., $\left<X^n\right>$, easily. 

Now let's calculate the generating function for the probability distributions \eqref{67}, \eqref{f-} and \eqref{f+}. By Eq.~\eqref{moment} we get
\begin{equation}\label{M-}
 M_-(s)= \frac{\exp[\frac{\sqrt{2} \sqrt{-s} \left(2 \phi _++\phi _--\phi _0\right)}{h }]-\exp[\frac{\sqrt{2} \sqrt{-s} \left(\phi _-+\phi _0\right)}{h }] }{\exp[ \frac{2 \sqrt{2} \sqrt{-s} \phi _+}{h }] -\exp[ \frac{2 \sqrt{2} \sqrt{-s} \phi _-}{h }] },
\end{equation}\\
\begin{equation}\label{M+}
  M_+(s)= \frac{\exp[\frac{\sqrt{2} \sqrt{-s} \left(2 \phi _-+\phi _+-\phi _0\right)}{h }] -\exp[ \frac{\sqrt{2} \sqrt{-s} \left(\phi _++\phi _0\right)}{h }] }{\exp[ \frac{2 \sqrt{2} \sqrt{-s} \phi _-}{h }] -\exp [\frac{2 \sqrt{2} \sqrt{-s} \phi _+}{h }] },
\end{equation}\\
\begin{equation}\label{Ms}
M(s)=M_+(s)+M_-(s)=\frac{\exp[\frac{\sqrt{2} \sqrt{-s} \left(\phi _-+\phi _+-\phi_0\right)}{h }]+ \exp[ \frac{\sqrt{2} \sqrt{-s} \phi _0}{h }] }{\exp[ \frac{\sqrt{2} \sqrt{-s} \phi _-}{h }]+\exp[\frac{\sqrt{2} \sqrt{-s} \phi _+}{h }] },
\end{equation}
where $M_\pm(s)$ are the generating function for the moments of $f_\pm$ and $M(s)$ is that of the total probability distribution function of hitting the barriers. By using Eq.~\eqref{Ms} we get
\begin{equation}
    \mu \left<F(\mathcal{N})\right>=\mu -\mu  \cosh \Big(\frac{\sqrt{\frac{3}{2}} \left(\phi _-+\phi _+-2\phi_0\right)}{h }\Big) \text{sech}\Big(\frac{\sqrt{\frac{3}{2}} \left(\phi _--\phi _+\right)}{h }\Big).
\end{equation}
By this expression as well as  $p_++p_-=1$ and Eq.~\eqref{63} one can see that
\begin{equation}
\begin{split}
  &p_+= \frac{\phi_0-\phi_-}{\phi _+-\phi _-} + \mu \frac{\cosh \Big(\frac{\sqrt{\frac{3}{2}} \left(\phi _-+\phi _+-2\phi_0\right)}{h }\Big) \text{sech}\Big(\frac{\sqrt{\frac{3}{2}} \left(\phi _--\phi _+\right)}{h }\Big)-1}{\phi _--\phi _+} +O(\mu^2),\\&
 p_-= \frac{\phi_+-\phi_0}{\phi _+-\phi _-} + \mu \frac{\cosh \Big(\frac{\sqrt{\frac{3}{2}} \left(\phi _-+\phi _+-2\phi_0\right)}{h}\Big) \text{sech}\Big(\frac{\sqrt{\frac{3}{2}} \left(\phi _--\phi _+\right)}{h }\Big)-1}{\phi _+-\phi _-}+O(\mu^2).
  \end{split}
\end{equation}

Now let's calculate $\left<F(\mathcal{N})W(\mathcal{N})\right>$ to read $\left<N\right>$ via Eq.~\eqref{64}. To this end at the zeroth order we write $\phi(N)-\phi_0=h W(N)+O(\mu)$. So up to zeroth order we can write
{{\begin{equation}\label{cond}
\begin{split}    \left<F(\mathcal{N})W(\mathcal{N})\right>&=\left<F(\mathcal{N})[\phi(\mathcal{N})-\phi_0]\right> \\
&=(\phi_+-\phi_0) p_+ \left<F(\mathcal{N})|\phi(\mathcal{N})=\phi_+\right> + (\phi_--\phi_0) p_- \left<F(\mathcal{N})|\phi(\mathcal{N})=\phi_-\right>,
    \end{split}
\end{equation}}}
where {{$\left<F(\mathcal{N})|\phi(\mathcal{N})=\phi_\pm\right>$}} are the conditional averages of $F(\mathcal{N})$ and are defined by 
{\begin{equation}
\left<F(\mathcal{N})|\phi(\mathcal{N})=\phi_\pm\right> \equiv \frac{1}{p_\pm} \int f_\pm(N) F(N)dN.
\end{equation}}
The conditional averages are calculated  as follows
{{\begin{equation}\label{cond1}
\begin{split}
&p_- \left<F(\mathcal{N})|\phi(\mathcal{N})=\phi_-\right>= \frac{\phi _+-\phi_0}{\phi _+-\phi _-} + \sinh \Big(\frac{\sqrt{6}(\phi _+-\phi_0)}{h }\Big) \text{csch}\Big(\frac{\sqrt{6} \left(\phi _--\phi _+\right)}{h }\Big),
\end{split}
\end{equation}
\begin{equation}\label{cond2}
\begin{split}
  &p_+  \left<F(\mathcal{N})|\phi(\mathcal{N})=\phi_+\right> = \frac{\phi _--\phi_0}{\phi _--\phi _+}-  \sinh \Big(\frac{\sqrt{6}(\phi _--\phi_0)}{h}\Big) \text{csch}\Big(\frac{\sqrt{6} \left(\phi _--\phi _+\right)}{h }\Big).
  \end{split}
\end{equation}}}
Substituting these   expressions into Eq.~\eqref{64} we find the following expression for the  mean number of $e$-folds:
 \begin{equation}
 \begin{split}
    \left<\mathcal{N}\right>&=-\frac{(\phi _--\phi_0)(\phi _+-\phi_0)}{h ^2}+\\&\frac{\mu}{h^2}  \Big[\left(\phi _+-\phi _-\right) \sinh \Big(\frac{\sqrt{\frac{3}{2}} \left(\phi _-+\phi _+-2\phi_0\right)}{h }\Big) \text{csch}\Big(\frac{\sqrt{\frac{3}{2}} \left(\phi _--\phi _+\right)}{h }\Big)+\phi _-+\phi _+-2\phi_0\Big]+O(\mu^2).\nonumber
    \end{split}
 \end{equation}
One can do the same procedure to get $\left<\mathcal{N}^2\right>$ up to $O(\mu)$, but because this expression is very complicated we avoid presenting it here.

Finally, we show the results of calculation of moments in the pure Brownian motion regime using the generating functions. One can easily show that
\begin{equation}
    \lim_{s\rightarrow0}M_\pm(s)=p_{\pm}=\mp\frac{\phi_\mp-\phi_0}{\phi_+-\phi_-},
\end{equation}
which is consistent with what we expect from the stochastic calculus. The other formulas one can obtain are 
\begin{equation}
    \left<\mathcal{N}\right>=\lim_{s\rightarrow0}M'(s)=-\frac{\left(\phi _--\phi _0\right) \left(\phi _+-\phi _0\right)}{h ^2},
\end{equation}
and
\begin{equation}
      \left<\mathcal{N}^2\right>=\lim_{s\rightarrow0}M''(s)= -\frac{(\phi _--\phi_0)( \phi _+-\phi_0)}{3 h^4} \big[(\phi _--\phi_0)^2-3 (\phi _+-\phi_0)(\phi _--\phi_0)+(\phi _+-\phi_0)^2\big].
\end{equation}

It is interesting to note that Eqs.~\eqref{M+} and \eqref{M-} can be used to obtain the results we derived in section~\ref{large kappa} by a simple differentiation: 
\begin{equation}
    \left<\mathcal{N}_+\right>= \lim_{s\rightarrow0}\frac{M_+'(s)}{p_+}=\frac{\left(\phi _+-\phi _0\right) \left(\phi _+-2 \phi _-+\phi _0\right)}{3 h ^2},
\end{equation}
and 
\begin{equation}
    \left<\mathcal{N}_-\right>= \lim_{s\rightarrow0}\frac{M_-'(s)}{p_-}=\frac{\left(\phi _--\phi _0\right) \left(\phi _--2 \phi _++\phi _0\right)}{3 h ^2}.
\end{equation}
Also we have
\begin{equation}
   \left<\mathcal{N}_+^2\right>= \lim_{s\rightarrow0}\frac{M_+''(s)}{p_+}=-\frac{\left(2 \phi _--\phi _+-\phi _0\right)}{{45 h^4}} \left[\phi _+-\phi _0\right) \left(4 \phi _-^2+\left(6 \phi _0-14 \phi _+\right) \phi _-+7 \phi _+^2-3 \phi _0^2\right],
\end{equation}
\begin{equation}
   \left<\mathcal{N}_-^2\right>= \lim_{s\rightarrow0}\frac{M_-''(s)}{p_-}= -\frac{\left(2 \phi _+-\phi_ --\phi _0\right)}{{45 h^4}} \left(\phi _--\phi _0\right) \left[4 \phi _+^2+\left(6 \phi _0-14 \phi _-\right) \phi _++7 \phi _-^2-3 \phi _0^2\right].
\end{equation}
Similarly one can calculate $\left<\mathcal{N}^n\right>$ for an arbitrary positive integer by the stochastic calculus (a la section~\ref{large kappa}) and the moment generating function approach.  Obviously, the two methods give the same result. However as we see the latter is much easier and faster to obtain the results.

\end{document}